\documentclass[journal]{IEEEtran}

\ifCLASSINFOpdf
\else
 \fi
\usepackage{amsmath,amssymb,enumerate,esint}
\usepackage{graphicx}
\usepackage{xcolor}
\usepackage{placeins}
\usepackage{float,dutchcal}
\usepackage{hyperref}
\usepackage{tabularx,colortbl}
\newcommand{\abs}[1]{\left\vert{#1}\right\vert}
\usepackage{mathtools}

\newcommand{\erfc}[1]{\mbox{erfc}(#1)}

\begin{document}
%
\title{Exact Transient Analysis of a Biconical Antenna of Arbitrary Length and Cone Angle}
%
%
%

\author{Ramakrishna Janaswamy,~\IEEEmembership{Fellow,~IEEE}
\thanks{R. Janaswamy is with the Department
of Electrical and Computer Engineering, University of Massachusetts, Amherst,
MA, 01003 USA. E-mail: janaswam@umass.edu. Submitted January 09, 2023, Accepted May 04, 2023.}
}

%
%

\markboth{to Appear in IEEE Transactions on Antennas and Propagation}
{RJ\MakeLowercase{\textit{et al.}}: Exact Transient Analysis of Biconical Antenna of Arbitrary Length and Cone Angle}
%



\maketitle

\begin{abstract}
Starting from a new factorization of the frequency-domain modal expansion coefficients, we generate analytical expressions in time-domain for the various quantities of interest of a biconical antenna of arbitrary length and cone angle. In particular, expressions are generated for the impulse response of feed current and the radiated fields as well as the transient effective length. Comparison with measured results available in the literature is carried out for the antenna effective length. The response due to any other feed voltage pulse can be obtained in terms of the impulse response via convolution. Results are presented for a Gaussian and a fast-rising feed voltage pulse. For the fast rising feed voltage pulse, the transient radiated energy density is shown to decay less rapidly than as inverse square distance over significant distances from the antenna. The expressions presented in the paper are all general and exact and not subject to any analytical approximations. 
\end{abstract}


%
\IEEEpeerreviewmaketitle

\section{Introduction}

\IEEEPARstart{T}{he} biconical antenna, see Fig.~\ref{fg:Biconical}, is an ultra-wideband (UWB) antenna \cite {Samaddar1997} that is capable of radiating very narrow pulses. The antenna was previously analyzed extensively for continuous wave (CW) operation by Schelkunoff \cite{Schelk}, \cite{Schelkunoff1952}, and Tai \cite{Tai} and the various analyses are summarized in \cite{CollZuck} and \cite{Jasik}. Approximate frequency-domain analyses were performed in \cite{papas1} and \cite{papas2} assuming that the field within the circumscribing sphere is comprised of only  the transverse electromagnetic wave. Analysis for non-equal cone-angles and a more elaborate analysis involving other electromagnetic metrics of interest were carried out in \cite{samaddar} and \cite{janaswamy2022}, respectively. Owing to its superior bandwidth properties, new variants in geometry are still being tried out for the biconical antenna within the UWB community \cite{McDonald}.

Time-domain studies pertaining to the biconical antennas are, however, few. This is despite the unmet demand triggered by UWB applications \cite{Reed}, \cite{Shlivinski}. Transient response of a wide-angle biconical antenna was studied in \cite{Harrison} starting from the approximate frequency-domain model of \cite{papas1} and \cite{papas2} and by employing the inverse Fourier transform. Electric Field Integral Equation (EFIE) in the frequency domain was utilized in \cite{Ghosh} in conjunction with an inverse Fast-Fourier-Transform (FFT) algorithm to generate the  pulse response of the biconical antenna. It may be mentioned parenthetically that the Fourier inversion based techniques operating on band-limited signals will necessarily result in non-casual signals. 

In contrast to these earlier approximate studies, we undertake here an exact analysis of the biconical antenna starting from a spherical harmonic expansion and by studying the analytical properties of the various functions in the complex frequency domain. This will lead directly to transient characteristics of the antenna when Laplace transform technique is employed, \`{a} la Singularity Expansion Method \cite{CEBaum}. An advantage of using the Laplace transform technique analytically as opposed to performing a numerical FFT on a band-limited frequency domain formulation is that it will properly result in a causal waveform. {In addition, the rigorous analytical results developed here will render the antenna as a much-needed benchmark case for validating various time-domain numerical simulation approaches. In this connection, it may be mentioned that even in the frequency domain where causality issues do not crop-up, the various numerical software tools are in disagreement with each other \cite{VmRJ2022} because they are subject to yet unresolvable errors arising from (i) improper geometric discretization, or (ii) improper computational domain discretization, or (iii) improper domain termination, or (iv) their inability to enforce auxiliary physical boundary conditions such as the edge condition, the tip condition, etc., or (v) lack of convergence, or (vi) combination of all of these. The solution developed here, on the other hand, is immune to such errors and is capable of incorporating all physical phenomena automatically within the formulation. Furthermore, it  provides insights on the time-domain performance of the antenna that are not possible to obtain with current numerical simulation approaches.}

Of particular mention here are the following new contributions: (i) the transient analysis carried out in the paper is exact, (ii) the spherical harmonic expansion is cast in a particular form that facilitates the expansion coefficients to be decomposed into a factor dependent on cone angle only and a factor dependent on the electrical length only so that the singularity expansion procedure can be carried out analytically, (iii) new expressions are derived for the impulse response of feed-current, the impulse response of radiated electric field, (iv) an expression is provided for the receive antenna effective length in time-domain, (v) pulse responses are obtained for commonly used input voltage excitations, and lastly, (vi) comparison is made for the time-domain effective length with the measured results available in \cite{LiculPhD}. 

In the next section we will initially provide field equations in the frequency domain and then obtain the desired time-domain responses by performing an inverse Laplace transformation analytically. More specifically, the initial formulation in the real frequency domain leads directly to a formulation in the complex wavenumber domain, which can then be transitioned to the complex frequency domain. 
\section{Theory and Results}
\subsection{Field Equations in the Frequency Domain}

\begin{figure}[htb]
\centerline{\scalebox{0.5}{\includegraphics{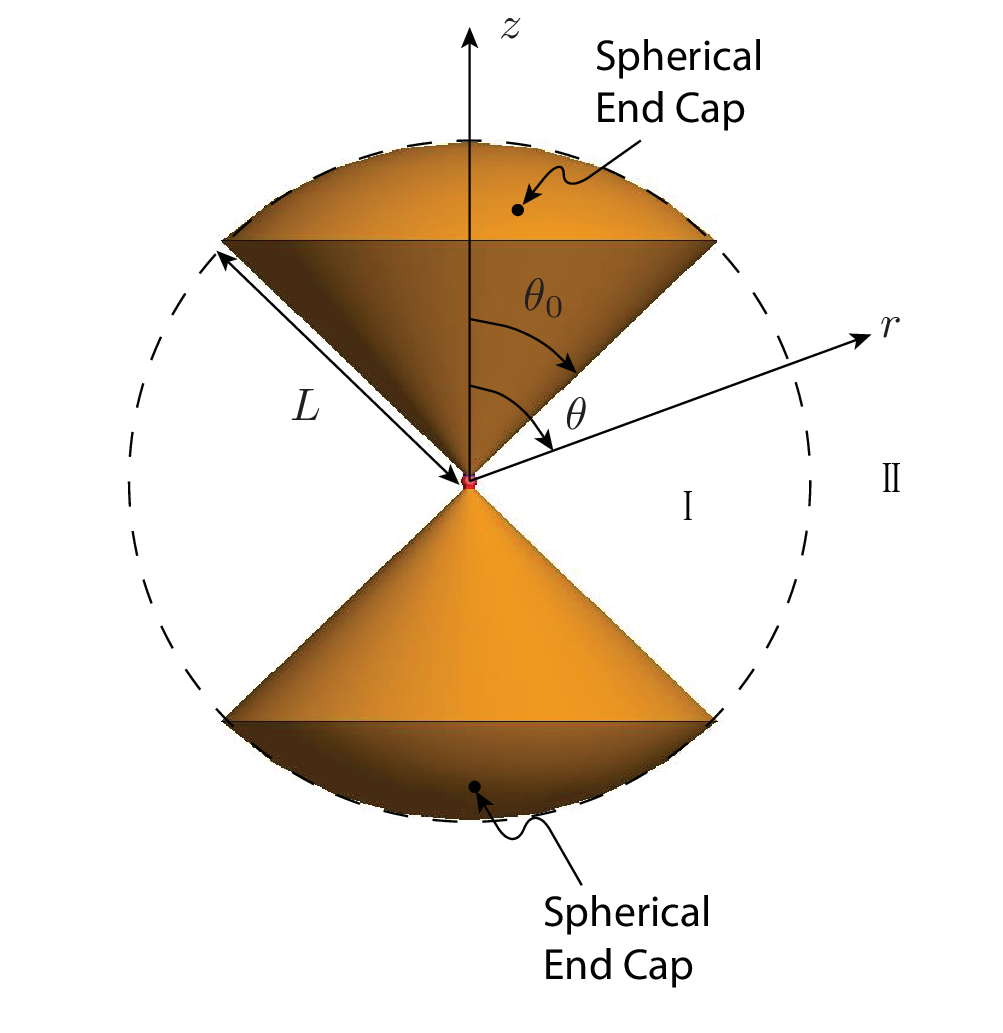}}}
\begin{center}
\caption{Symmetrical biconical antenna having length $L$ and half-angle $\theta_0$. The cones are sealed off with perfectly conducting spherical caps.}
\label{fg:Biconical}
\end{center}
\end{figure}

A symmetric, coaxial biconical antenna having identical, perfectly conducting (PEC) cones with half angles $\theta_0$ and axes aligned along the $z$-axis is shown in Fig.~\ref{fg:Biconical}. The entire antenna is contained with a radis $L$, which we label as the length of the antenna. The azimuth angle not shown in the figure is denoted by $\phi$. The region between the two cones (region-I) is $\theta_0\le \theta\le \pi-\theta_0,\ 0\le\phi\le2\pi,\ 0<r\le L$. The region exterior to $r=L$ is labelled as region-II. The cones are terminated in spherical PEC caps\footnote{A practical biconical antenna may have flat end-caps or it may be hollow. Spherical end-caps are used here to permit an exact analytical solution. The difference between a flat-capped and a spherical-capped antenna is expected to result in second order effects. In Figure~\ref{fg:CompExp} we show favorable comparison between the present results employing spherical caps and experimental results of \cite{LiculPhD} performed on a hollow biconical antenna.} at the two ends. An infinitesimal source of frequency-dependent voltage $V_g(\omega)$ is applied at the apex of the antenna, where an $e^{j\omega t}$ time convention at the angular frequency $\omega$ is assumed. The medium exterior to the cones is assumed to be vacuum with electrical parameters $(\epsilon_0,\mu_0)$.

We assume operation in a balanced mode where the transverse electric field near the apex is an even function of polar angle $\theta$ about the $\theta = \pi/2$ plane. The only non-zero components of field in such a case are $E_r,E_\theta,$ and $H_\phi$, which are all azimuthally invariant. Therefore the field is comprised of lower order transverse magnetic (TM$_r$) modes, which have no azimuthal variation. In addition to these modes, the structure supports a transverse electromagnetic (TEM) mode within the circumscribed sphere (that is, in the region $0\le r\le L$). As stated previously, the analysis performed in \cite{papas1} and \cite{papas2} are approximate and subject to the assumption that only a TEM mode exists within the circumscribing sphere. However, there is evidence that the non-TEM modes that exist in region-I can play a significant role \cite{Blume} in the impedance as well as in the radiative properties of the antenna. Hence we do not make such dubious assumptions here, but will, instead, expand the fields in both regions I and II in terms of complete sets.  

The fields in region-I are expanded as  
\begin{eqnarray}
 rE_r&=&\frac{V_g(\omega)LA_0(k)}{r}\sum\limits_{\nu_n}a_{\nu_n}(k)\widehat{J}_{\nu_n}(kr)M_{\nu_n}(\theta),\label{eq:rEr1}\\
 rE_\theta&=&V_g(\omega)kLA_0(k)\sum_{\nu_n}\frac{a_{\nu_n}(k)\widehat{J}^{\prime}_{\nu_n}(kr)M^\prime_{\nu_n}(\theta)}{\nu_n(\nu_n+1)}\nonumber\\
 &&+ rE_{\theta 0},\label{eq:rEth1}\\
 rH_\phi&=&-j\frac{V_g(\omega)kLA_0(k)}{\eta}\sum\limits_{\nu_n}\frac{a_{\nu_n}(k)\widehat{J}_{\nu_n}(kr)M^\prime_{\nu_n}(\theta)}{\nu_n(\nu_n+1)}\nonumber\\
 &&+\ rH_{\phi 0},\label{eq:rHph1}
\end{eqnarray}
where $\widehat{J}_{\nu_n}(\cdot)$ is the spherical Riccati-Bessel function of order, $\nu_n$, $M_{\nu_n}(\theta) = [P_{\nu_n}(\cos\theta)-P_{\nu_n}(-\cos\theta)]/2$ is an odd function of $\theta$ about $\theta=\pi/2$, $P_\nu(\cdot)$ is the Legendre function of  degree $\nu$ (possibly non-integer), $\eta = \sqrt{\mu_0/\epsilon_0}$ is the intrinsic impedance of free-space, $k = \omega\sqrt{\mu_0\epsilon_0} = \omega/c$ is the wavenumber in free-space at the operating frequency $\omega$, and $c$ is the speed of light in free-space. A prime on Bessel functions denotes derivative with respect to its entire argument. The spherical TEM fields $(E_{\theta 0}, H_{\phi 0})$ include both outgoing and incoming waves and are given by  
\begin{eqnarray}
E_{\theta 0}&=& \frac{V_g(\omega)A_0(k)}{r\sin\theta}\left(e^{-jkr}+\Gamma_{\rm in}(k)e^{jkr}\right),\label{eq:ETEM}\\
H_{\phi 0}&=&\frac{V_g(\omega)A_0(k)}{r\eta\sin\theta}\left(e^{-jkr}-\Gamma_{\rm in}(k)e^{jkr}\right).\label{eq:HTEM}
\end{eqnarray}
For ease of notation, arguments of fields and voltages showing explicit dependence on $(r,\theta)$ and/or $\omega$ are suppressed when there is no cause of confusion. The quantity $\Gamma_{\rm in}(k)$ is the frequency-dependent reflection coefficient of the TEM wave and the factor $A_0(k)$ is unitless. The form of mode expansion coefficients $a_{\nu_n}$ used here is different from that assumed in previous works and in \cite{janaswamy2022}. These coefficients together with the constants appearing in front of the summation sign in (\ref{eq:rEr1}) and the superposition of forward and backward waves in (\ref{eq:ETEM}) are written in a form to facilitate convenient factorizations of the frequency and angular-dependent terms (that is, those related to $\theta_0$) in the ensuing analysis. These nuances are less important if one remains entirely in the frequency domain. However, they will play an important role in the study of singularities in the complex frequency domain and in the derivation of various time-domain waveforms.

The voltage across the feed-gap can be expressed entirely in terms of the TEM electric field as $V(0;k) = \lim_{r\to 0}\int_{\theta_0}^{\pi-\theta_0}rE_{\theta 0}\,d\theta = V_g(\omega)A_0(k)(1+\Gamma_{\rm in}(k))2\ln[\cot(\theta_0/2)]$. In order for this to be equal to the assumed gap voltage of $V_g(\omega)$ we need $2A_0(k)(1+\Gamma_{\rm in}(k))\ln[\cot(\theta_0/2)] = 1$ or equivalently,
\begin{equation} A_0(k) = \frac{C_0}{1+\Gamma_{\rm in}(k)}\label{eq:A0},\end{equation} where $C_0 = \eta Y_0/2\pi=1/2\ln[\cot(\theta_0)]$ and $Z_0 = Y_0^{-1}$ is the characteristic impedance of an infinite biconical transmission line. The current, $I(0;k)$, crossing the feed-gap in the positive $z$-direction can be expressed entirely in terms of the TEM magnetic field and equals \begin{eqnarray}
I(0;k) &=&\lim_{r\to 0\atop \theta=\pi/2}\oint rH_{\phi 0}\,d\phi = 2\pi \frac{V_g(\omega)}{\eta}A_0(k)(1-\Gamma_{\rm in}(k))\nonumber\\
&=&V_g(\omega)Y_0\, \frac{1-\Gamma_{\rm in}(k)}{1+\Gamma_{\rm in}(k)} = V_g(\omega)Y_{\rm in}(\omega),\label{eq:Ifeed} \end{eqnarray}
where $Y_{\rm in}$ is the antenna input admittance. Note that the feed current is, in general, frequency dependent even when the voltage $V_g(\omega)$ applied across the gap is frequency independent. The frequency dependence of $I(0;k)$ arises from reflections taking place at the ends of the antenna, which will vary with $L$ and $\theta_0$ in addition to the frequency. The unknowns contained in the interior fields are $\nu_n, a_{\nu_n}$ and $A_0(k)$, or equivalently, $\Gamma_{\rm in}(k)$. 

The eigenvalues $\nu_n$, $n=1,2,\ldots$ are determined by imposing the boundary conditions at $\theta=\theta_0$ as discussed in detail in \cite{janaswamy2022} and will not be repeated here. Suffice it to say that they depend on the cone angle $\theta_0$ and  accurate asymptotic formulas for determining them are 
\begin{equation}
\nu_n \sim \sqrt{n^2\alpha_0^2-0.25} - 0.5,\ \frac{\partial\nu_n}{\partial\theta_0}\sim \frac{n^2\alpha_0^3}{\pi(\nu_n+0.5)},\ \nu_n\gg 1,\label{eq:nudnu}
\end{equation} 
where $\alpha_0 = (0.5-\theta_0/\pi)^{-1}$. 

The fields in the region $L\le r<\infty$ (region-II) are expanded in terms of TM$_r$ modes as  
\begin{eqnarray}
rE_r&=&\frac{V_g(\omega)LA_0(k)}{r}\sum\limits_{\ell=1,3,\dots}^\infty j^{-\ell}\ell(\ell+1)b_{\ell}(k)\widehat{H}_\ell^{(2)}(kr)\nonumber\\
&&\hskip 1in \times\, P_\ell(\cos\theta),\label{eq:rEr2}\\
rE_\theta&=&V_g(\omega)kLA_0(k)\sum\limits_{\ell=1,3,\dots}^\infty j^{-\ell}b_{\ell}(k)\widehat{H}_\ell^{(2)\prime}(kr)\nonumber\\
&&\hskip 1in \times\, P_\ell^1(\cos\theta),\label{eq:rEth2}\\
rH_\phi&=&-j\frac{V_g(\omega)kLA_0(k)}{\eta}\sum\limits_{\ell=1,3,\dots}^\infty j^{-\ell}b_{\ell}(k)\widehat{H}_\ell^{(2)}(kr)\nonumber\\
&&\hskip 1in \times\, P_\ell^1(\cos\theta),\label{eq:rHph2}
\end{eqnarray}
where $P_\ell^1(\cos\theta)=dP_\ell(\cos\theta)/d\theta$ and $\widehat{H}_{\ell}^{(2)}(\cdot)$ is the spherical Riccati-Hankel function of the second  kind of order $\ell$. These field expansions are consistent with the Bouwkamp and Casimir's theorem \cite[p. 1-27:1-29]{RJ2020} in $r\ge L$. The coefficients of expansion $b_\ell(k)$ are, in general, complex-valued and depend on the cone angle $\theta_0$ as well as on the electrical length $kL$ of the antenna. 

In subsequent analysis we shall use the revised expansion coefficients, $c_\ell$ (real-valued), introduced in equation (\ref{eq:coeffcl}) of Appendix~\ref{sc:App1} in place of $b_\ell(k)$. The linear system of equations needed to solve for these coefficients is given in (\ref{eq:cell2Eqn}). There, the expansion coefficients are factorized into separable functions of angle and frequency. Given the cone angle $\theta_0$, the linear system (\ref{eq:cell2Eqn}) need only be solved once completely and is {\em independent of antenna electrical length}. Formulation in terms of the revised coefficients offers this distinct advantage over the previous formulations of the biconical antenna. Another important advantage is that the revised form permits analytical determination of Laplace inverse that is needed for time-domain representations. 

Letting
\begin{eqnarray}
u_\ell^E(kr) &=& j^{-\ell}\widehat{H}_\ell^{(2)\prime}(kr)e^{jkr}\to 1, \mbox{ as } kr\to \infty,\label{eq:uEl}\\
u_\ell^H(kr) &=& j^{-\ell-1}\widehat{H}_\ell^{(2)}(kr)e^{jkr}\to 1, \mbox{ as } kr\to \infty,\label{eq:uHl}
\end{eqnarray}
the expressions for the radiated electric and magnetic fields for any $r\ge L$ can be re-cast using the expression of $\Gamma_{\rm in}$ given in (\ref{eq:Gamin}) as
\begin{eqnarray}
rE_\theta(r,\theta;k) &=&-V_g(\omega)C_0e^{-jkr}\,\frac{F_e({\bf r};\zeta)}{F(\zeta)},\label{eq:rErad}\\
rH_\phi(r,\theta;k) &=&-\frac{V_g(\omega)C_0}{\eta}e^{-jkr}\,\frac{F_h({\bf r};\zeta)}{F(\zeta)},\label{eq:rHrad}
\end{eqnarray}
where $\zeta = kL=\omega t_0$ is the normalized frequency. The pattern functions $F_e({\bf r};\zeta)$ and $F_h({\bf r};\zeta)$ associated with the electric and magnetic fields are 
\begin{equation}
F_{\{{e\atop h}\}}({\bf r};\zeta) = \sum\limits_{\ell=1,3}^\infty \frac{j^\ell u_\ell^{\{{E\atop H}\}}(kr)c_\ell P_\ell^1(\cos\theta)}{\widehat{H}_\ell^{(2)}(\zeta)+j\widehat{H}_\ell^{(2)\prime}(\zeta)}\label{eq:Feh}
\end{equation}
and the function $F(\zeta) = [1+\Gamma_{\rm in}(\zeta)]\,e^{j\zeta}/2j $ equals
\begin{equation}
F(\zeta) =\sum\limits_{\ell=1,3}^\infty w_\ell\frac{\sin\zeta\, \widehat{H}_\ell^{(2)}(\zeta)+\cos\zeta\,\widehat{H}^{(2)\prime}_\ell(\zeta)}{\widehat{H}_\ell^{(2)}(\zeta)+j\widehat{H}^{(2)\prime}_\ell(\zeta)}\label{eq:Fz}
\end{equation}
where $w_\ell$ is defined in (\ref{eq:check1}). 
The relation between $A_0(\zeta)$ and $F(\zeta)$ is \begin{equation}
A_0(\zeta) =  \frac{C_0e^{j(\zeta-\pi/2)}}{2F(\zeta)}.\label{eq:FtoA0}
\end{equation}
Equation (\ref{eq:Gamlmts}) implies the following asymptotic results
\begin{equation}
F(0) = -j;\ F(\zeta\to\infty) \sim -0.5j\,e^{j\zeta}.\label{eq:Fzlmts} \end{equation}
The far-zone electric field (respectively, magnetic field) is obtained by simply replacing $u_\ell^E(kr)$ (respectively, $u_\ell^H(kr)$) with unity in (\ref{eq:Feh}). 

We define a fourth function $F_{el}(\zeta) = [1-\Gamma_{\rm in}(\zeta)]e^{j\zeta}/2$, which upon using (\ref{eq:Gamin}), equals 
\begin{equation}
F_{el}(\zeta) =\sum\limits_{\ell=1,3}^\infty w_\ell\frac{\cos\zeta\, \widehat{H}_\ell^{(2)}(\zeta)-\sin\zeta\,\widehat{H}^{(2)\prime}_\ell(\zeta)}{\widehat{H}_\ell^{(2)}(\zeta)+j\widehat{H}^{(2)\prime}_\ell(\zeta)}.\label{eq:Felz}
\end{equation}
In terms of the functions $F_{el}(\zeta)$ and $F(\zeta)$, the feed current can be expressed as 
\begin{equation}
I(0;\zeta) = V_g(\omega)Y_0\left[\frac{1-\Gamma_{\rm in}(\zeta)}{1+\Gamma_{\rm in}(\zeta)}\right] = -jV_g(\omega)Y_0\frac{F_{el}(\zeta)}{F(\zeta)}.\label{eq:I0zeta}
\end{equation}

The complex-valued {\em vector effective length} of the antenna, ${\bf h}_e^{Rx}$, in a receive mode is related to the antenna radiated field, ${\bf E}_a$, and the feed current and defined as \cite[p. 105]{CollZuck}
\begin{equation}
{\bf h}_e^{Rx} = \frac{4\pi r{\bf E}_ae^{jkr}}{jk\eta I(0;k)}.\quad\mbox{[m]}\label{eq:vecefflen} 
\end{equation}
In terms of the effective length and the incident electric field, ${\bf E}^i$, the open-circuited received voltage, $V_{oc}$, is 
\begin{equation}
V_{oc} = {\bf h}_e^{Rx}\cdot{\bf E}^i.\label{eq:Voceq}
\end{equation}
The $\hat{\bf\theta}$-component of the effective length here is equal to 
\begin{equation}
h^{Rx}_e(\theta,r;\zeta) = \frac{4\pi rE_\theta({\bf r};\zeta) e^{jkr}}{j\zeta\eta I(0;\zeta)}L
=\frac{-2LF_e({\bf r};\zeta)}{\zeta F_{el}(\zeta)}.\label{eq:heff}
\end{equation}
Strictly speaking, the effective length must be independent of $r$ because it is usually defined using the far-zone radiated fields ${\bf E}_a$ (note that the factor $u_\ell^E(kr)\to 1$ in the far-zone). However, we have retained the general expression of radiated field for arbitrary distances in (\ref{eq:heff}) so that the formula remains valid when measurements of effective length are conducted in the {\em near-field} of the antenna. 

It is interesting to note upon using (\ref{eq:check1}) that the functions $F(\zeta)$ and $F_{el}(\zeta)$ satisfy
\begin{equation}
F_{el}(\zeta) + jF(\zeta) = e^{j\zeta}\label{eq:FFelz}
\end{equation}
from which one can conclude upon using (\ref{eq:Fzlmts}) and (\ref{eq:Gamlmts}) that 
\begin{equation}
F_{el}(\zeta\to 0) = O(\zeta);\ F_{el}(\zeta\to \infty) \sim 0.5\,e^{j\zeta}.\label{eq:Fellmts}
\end{equation}
Note that the various functions of $\zeta$ involved in $h_e^{Rx}$, $I(0,\zeta)$ and $E_\theta$ can be analytically continued into the complex $\zeta$ plane by considering the medium exterior to the antenna to be vanishingly lossy, which is equivalent to replacing the real-valued wavenumber $k$ with the complex wavenumber $k(1-j\epsilon_\ell)$, where $\epsilon_\ell>0$ arises from the slight loss. The resulting complex-valued $\zeta$ is directly related to the complex frequency variable $s$ that is used in the ensuing analysis. All functions involved in the various expressions above can be analytically continued into the complex $\zeta$-plane. 
   
It is also possible to define an antenna transfer function directly between the radiated electric field and the input voltage after removing the free-space phase and spreading factor as 
\begin{equation}
T_0^e({\bf r};k) = \frac{4\pi rE_\theta({\bf r};k)}{V_g(\omega)} e^{jkr}
=-4\pi C_0\frac{F_e({\bf r};\zeta)}{F(\zeta)}.\label{eq:T0ef}
\end{equation}
This transfer function too will be dependent on range, observation angles and frequency of operation. A transfer function of this kind is useful in optimizing the source voltage waveform to achieve a desired radiative property from the antenna. The relation between $h_e^{Rx}$ and $T_0^e$ is $h_e^{Rx} = T_0^e/jk\eta Y_{\rm in}$ from which it is clear that the effective length accentuates lower frequencies and deemphasizes higher frequencies present in $T_0^e$. The effective length serves as a modified transfer function between the antenna radiated field and the antenna feed current and is frequently used in the extraction of time-domain characteristics from frequency domain measurements \cite{Reed} as well as in the optimization of pulse waveforms generated by antennas \cite{pozar1}. 
\subsection{Time-Domain Characteristics}
The time-domain waveform corresponding to a complex frequency-domain quantity is obtained by taking the inverse Laplace transform. Let $t_0 = L/c$ be the transit time of a wave as it traverses the antenna length. For generating the impulse response of the feed current, the input feed voltage is taken to be $v_g(t) = V_0\delta(t-\epsilon t_0)$, [Vsec$^{-1}$], where $\epsilon>0$ is a vanishingly small number. The Laplace transform of this impulse function in the transform variable $s$ is $V_0e^{-\epsilon st_0}\,\to V_0$ as $\epsilon\to 0$. 

The entire antenna may be regarded as a linear system having a single input and multiple outputs. In the analysis to follow we will regard the applied input voltage as the {\em cause} in the linear system. Because the antenna does not behave as a purely resistive system at all frequencies, but rather as an inertial system with reactive elements too, the outputs associated with several observables such as the feed current, the vector effective length, the vector radiated fields, etc., will be subject to a certain time delay before they respond. These entities all constitute the {\em effects} of the linear system. Keeping causality in mind, we are thus interested in the various responses for $t\lessgtr \epsilon t_0$, but not precisely at $t = \epsilon t_0$.  

The impulse response of the feed current can be written in terms of the inverse Laplace transform as 
\begin{eqnarray}
t_0i_\delta(t) &=& \frac{t_0}{2\pi j}\int\limits_{\Gamma}I(0;\zeta)e^{st}\,ds = \frac{1}{2\pi j}\int\limits_{\Gamma_q}I(0;\zeta)e^{q\tau}\,dq\nonumber\\
&=&-\frac{Y_0V_0}{2\pi}\int\limits_{\Gamma_q}\left.\frac{F_{el}(\zeta)}{F(\zeta)}\right\vert_{\zeta =-j q}e^{q(\tau-\epsilon)}\,dq,\label{eq:IRcur1}
\end{eqnarray}
where $\tau = t/t_0$ is the normalized time, $q = st_0$ is the normalized complex frequency corresponding to the actual complex frequency $s=jkc=j\zeta/t_0=q/t_0$, $\Gamma$ is the Bromwich contour \cite[p. 275]{Ablowitz} as shown in Fig.~\ref{fg:Brmwch} and $\Gamma_q$ is the corresponding contour in the $q$-plane. {In practice, the Bromwich contour is closed by infinite semicircular arcs in the right half-plane (RHP) or the left half-plane (LHP) of the complex $s$-plane as shown in Fig.~\ref{fg:Brmwch}. This is done to capture any singularities in the respective half-planes while evaluating the inversion integral. Causality requires that the integral (\ref{eq:IRcur1}) for impulse response be zero for $\tau <\epsilon$. For $\tau<\epsilon$, the Bromwich contour is closed by $S_\infty^-$. If $I(0;\zeta)$ is analytic the right half of the $q$-plane, then the inversion integral vanishes by virtue of Cauchy's theorem \cite[p. 83]{Ablowitz} and from the fact that $e^{q(\tau-\epsilon)}$ vanishes on $S_\infty^-$. Hence causality requires that $I(0;\zeta)$ be free from singularities to the right of $\Gamma_q$ in the complex $q$-plane. Moreover, due to the presence of the exponential factor $e^{q(\tau-\epsilon)}$, the integrand will vanish on the semicircular arc $S_\infty^+$ for arbitrary $\tau>\epsilon$ even if the factor $\abs{\frac{F_{el}(\zeta)}{F(\zeta)}}\ne o(1)$ in the LHP as $\abs{\zeta}\to\infty$. By closing off the Bromwich contour by $S_\infty^+$ for $\tau>\epsilon$ and applying Cauchy's residue theorem \cite[p. 207]{Ablowitz}, the inversion integral will then become equal to the sum of residues at the pole singularities and sum of integrals around any branch cuts present in the LHP. If $\frac{F_{el}(\zeta)}{F(\zeta)}\equiv j$, then the inversion integral will still yield a value zero for $\tau>\epsilon$, but at the single instant $\tau = \epsilon$ it will yield the functional $V_0Y_0\delta(\tau-\epsilon)\to\infty$. In this connection it may be mentioned that the usual requirement that the Laplace transform behave as $\lim_{s\to\infty}\abs{\widetilde{F}(s)} = 0$ is only a sufficiency condition \cite[p. 88]{Debnath}, \cite[p. 1054]{deHoopbook} and that it is made to ensure that the Laplacian inversion integral 
\begin{equation}
f(t) = \frac{1}{2\pi j}\int\limits_{\Gamma}\widetilde{F}(s)e^{st}\,ds\label{eq:LTinv}
\end{equation}
will result in the function $f(t)$ that is continuous or piece-wise continuous in every finite interval $t\in(0,T), T>0$. 
\begin{figure}[htb]
\centerline{\scalebox{0.4}{\includegraphics{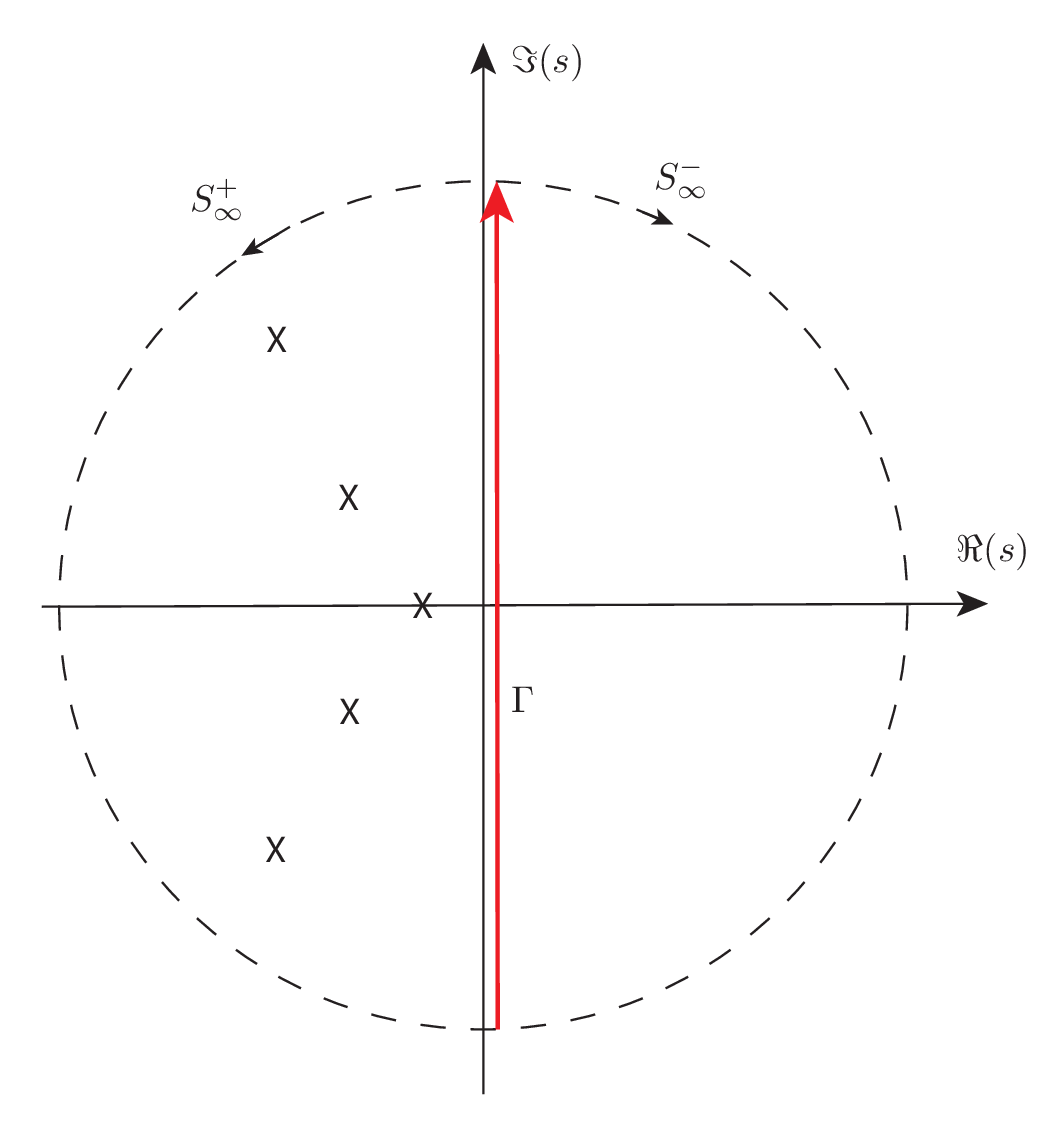}}}
\begin{center}
\caption{Bromwich contour $\Gamma\in(a-j\infty,a+j\infty)$ in the complex $s$-plane for some $a>\epsilon>0$. All singularities lie to the left of the contour $\Gamma$.}
\label{fg:Brmwch}
\end{center}
\end{figure} 

Availability here of expressions in the frequency domain showing explicit complex frequency dependences of various terms enables an analytical Laplace inversion not possible otherwise. For example, for the expression of input reflection coefficient given in (\ref{eq:Gamin}), the explicit frequency dependence is governed by the complex argument behavior of the spherical Riccati Hankel functions $\widehat{H}_\ell^{(2)}(\zeta)$ and $\widehat{H}^{(2)\prime}_\ell(\zeta)$. 

Before the inverse Laplace transform of various quantities is undertaken it is necessary to study the distribution of singularities present in the various integrands. The following observations can be made (recalling that $q =st_0= j\zeta$): 
\begin{enumerate}[(1)]
\item The functions $F(\zeta), F_e(\zeta), F_h(\zeta)$ and $F_{el}(\zeta)$ are all unitless. 
\item At the zeros of the function $f_\ell(\zeta) = \widehat{H}_\ell^{(2)}(\zeta)+j\widehat{H}_\ell^{(2)\prime}(\zeta)$ in the complex-$\zeta$ plane, the reflection coefficient $\Gamma_{\rm in}(\zeta)\to\infty$. However, the feed current reduces to $I(0;\zeta)=-V_0Y_0$ and remains finite there. Furthermore, the function $f_\ell(\zeta)$ appears in both the numerator and denominator (via the factor $F(\zeta)$) of $rE_\theta$ as well as of $h_\theta^{Rx}$, rendering both of them finite there. Hence zeros of  $f_\ell(\zeta)$ are not singularities of $I(0;\zeta)$ or $rE_\theta({\bf r};\zeta)$ or $h_e^{Rx}({\bf r};\zeta)$.  
\item Singularities of {\em both} $I(0;\zeta)$ and $rE_\theta({\bf r};\zeta)$ occur at the zeros of the function $1+\Gamma_{\rm in}(\zeta)$; in other words, at the zeros of the function $F(\zeta)$ given in (\ref{eq:Fz}). These are complex frequencies at which the feed-gap appears as a short circuit.  
\item Singularities of the effective length $h_e^{Rx}(\theta;\zeta) $ occur at the zeros of the function $F_{el}(\zeta)$ given in (\ref{eq:Felz}), which correspond to the zeros of the function $1-\Gamma_{\rm in}(\zeta)$. These are complex frequencies at which the feed gap appears as an open circuit. In the far-zone where $u_\ell^E(kr)\to 1$, there is no singularity of $h_e^{Rx}$ at $\zeta = 0$ despite the presence of the term $\zeta$ in the denominator of the right-hand-side of (\ref{eq:heff}). 
\item It will be clear in the ensuing that the singularities of $I(0;\zeta)$, $rE_\theta({\bf r};\zeta)$ and $h_e^{Rx}({\bf r};\zeta)$ are all simple poles in the complex $\zeta$-plane. In particular, there are no branch point singularities (this will be made clear in the next item), in compliance with the conjecture made by Baum \cite{CEBaum} that  the only type of singularities for a finite-sized perfectly conducting object are simple poles. 
\item The functions $F(\zeta)$ and $F_{el}(\zeta)$ are both in terms of single-valued functions; consequently they do not contain any branch points. Note that the spherical Riccati Hankel function $\widehat{H}_\ell^{(2)}(\zeta)$, which appears in $F(\zeta), F_{el}(\zeta)$ as well as in other parts of the expressions for $I(0;\zeta)$, $rE_\theta$ and $h_e^{Rx}$, is single-valued for integer $\ell$. This is clear from its series representation $\widehat{H}_\ell^{(2)}(\zeta) = e^{-j\zeta}j^{\ell+1}\sum\limits_{n=0}^\ell(\ell+n)!(2j\zeta)^{-n}/n!(\ell-n)!$. 
\item At dc $\zeta \to 0$ and from the small argument approximation of the spherical Riccati Hankel functions \cite[Appendix C]{RJ2020}, $\widehat{H}^{(2)}_\ell(\zeta)\sim j(2\ell)!(2\zeta)^{-\ell}/\ell!$, $\widehat{H}^{(2)\prime}_\ell(\zeta)/\widehat{H}^{(2)}_\ell(\zeta)\sim -\ell/\zeta$, it is easy to show that
\begin{eqnarray}
\Gamma_{\rm in}(\zeta)\sim 1 + O(\zeta)&\implies& I(0;\zeta)\sim O(\zeta)\label{eq:dcGamI0}\\
F_e(r=L,\theta;\zeta)&\sim& O(1)\label{eq:dc1Fe}\\
F_h(r=L,\theta;\zeta)&\sim& O(\zeta)\label{eq:dc1Fh}\\
F_e(r\to\infty,\theta;\zeta)&\sim& O(\zeta^2)\label{eq:dc2Fe}\\
F_h(r\to\infty,\theta;\zeta)&\sim& O(\zeta^2)\label{eq:dc2Fh}\\
 h_e^{Rx}(r\to\infty,\theta;\zeta)&\sim& \frac{-2LP_1^1(\cos\theta)c_1}{\sum\limits_{\ell=1,3}^\infty (1+\ell^{-1})w_\ell(\theta_0)}.\label{eq:dche}\end{eqnarray}
Thus different near-fields exhibit different dc behavior. However, the dc feed current and the dc far-zone radiated fields are identically zero as expected. 
\item At high frequencies, $\zeta\to\infty$ and
\begin{equation}
F_{\{{e\atop h}\}}(kr\gg 1,\theta;\zeta\to\infty) = O(e^{j\zeta}).
\end{equation}
Therefore, in view of (\ref{eq:Fzlmts}), we have in the far-zone region $kr\gg 1$ that 
\begin{equation}
E_\theta(\zeta\to\infty) = O(e^{-j\zeta r/L}) = H_\phi(\zeta\to\infty).\label{eq:EHHf}
\end{equation}
\item From (\ref{eq:Fzlmts}) and (\ref{eq:Fellmts}) 
it is clear that the factor
\begin{equation}
\frac{F_{el}(\zeta)}{F(\zeta)}\sim j \mbox{ as }\abs{\zeta}\to\infty.\label{eq:FeloverFinf}
\end{equation} 
\item Using the property $\widehat{H}_\ell^{(2)}(e^{-j\pi}\zeta^*) = (-1)^{\ell+1}\widehat{H}^{(1)}_\ell(\zeta^*)$ of the spherical Riccati Hankel function, where the superscript * denotes complex conjugation, it is easy to conclude that $F(-\zeta^*) = -F^*(\zeta)$ and $F_{el}(-\zeta^*) = -F_{el}^*(\zeta)$. Therefore, if $\zeta_0=\omega_0 +j\sigma_0$ is a zero of $F(\zeta)$, so is $-\zeta_0^*=-\omega_0+j\sigma_0$. Likewise, if $\zeta_p=\omega_p+j\sigma_p$ is a zero of $F_{el}(\zeta)$, so is $-\zeta_p^*=-\omega_p+j\sigma_p$. The corresponding zeros in the complex $q$-plane then appear as complex conjugate pairs since $q = j\zeta$. 
 \item Using the differential equation satisfied by the spherical Riccati Hankel function, it is straightforward to see that near a zero $\zeta_0$ the function $F(\zeta)$ behaves as  
 \begin{equation}
 F(\zeta)\sim \frac{e^{-j\zeta_0}}{\zeta_0^2}(\zeta-\zeta_0)G(\zeta_0),\label{eq:Fz0}
 \end{equation}
 where 
 \begin{eqnarray}
 G(\zeta) &=& \sum\limits_{\ell = 1,3}^\infty\ell(\ell+1)w_\ell\left[\frac{\widehat{H}_\ell^{(2)}(\zeta)}{\widehat{H}^{(2)}_\ell(\zeta)+j\widehat{H}^{(2)\prime}_\ell(\zeta)}\right]^2\nonumber\\
 &=&G^*(-\zeta^*).\label{eq:Gofz}
 \end{eqnarray}
 It can be verified by direct substitution that $G(\zeta_0)\ne 0.$ Hence the zero of $F(\zeta)$ is simple. Similarly, near a zero $\zeta_p$ of $F_{el}(\zeta)$
 \begin{equation}
 F_{el}(\zeta)\sim -j\frac{e^{-j\zeta_p}}{\zeta_p^2}(\zeta-\zeta_p)G(\zeta_p).\label{eq:Felzp}
 \end{equation}
It can again be verified that $G(\zeta_p)\ne 0$, implying that the zero of $F_{el}(\zeta)$ is also simple. 
\end{enumerate}
It is interesting to note that the function $G(\zeta)/\zeta^2$ that appears in (\ref{eq:Fz0}) and (\ref{eq:Felzp}) is equal to the Wronskian \cite[p. 75]{Kreyzig} $W[F_{el},F](\zeta) = F^\prime(\zeta)F_{el}(\zeta)-F_{el}^\prime(\zeta)F(\zeta)$ of $F(\zeta)$ and $F_{el}(\zeta)$. This can be verified by making use of (\ref{eq:FFelz}). For $\zeta \sim 0$, with $\hat{H}^{(2)\prime}_\ell(\zeta)/\hat{H}^{(2)}_\ell(\zeta) \sim -\ell/\zeta$, it can be shown that $\lim_{\zeta\to 0} [G(\zeta)/\zeta^2] = \sum\limits_{\ell}w_\ell(\ell+1)/\ell = 1+\sum\limits_\ell w_\ell/\ell\ne 0$. Hence $F_{el}(\zeta)$ and $F(\zeta)$ are linearly independent at the origin. From Abel's identity \cite[p. 75]{ince}\footnote{This assumes that the functions $F(\zeta)$ and $F_{el}(\zeta)$ satisfy the same second order linear differential equation in the independent variable $\zeta$.} one concludes that $F_{el}(\zeta)$ and $F(\zeta)$ are also linearly independent over an extended region $S$ containing the origin, implying that $G(\zeta)/\zeta^2\ne 0$ over $S$. This also proves that the zeros of $F(\zeta)$ and $F_{el}(\zeta)$ are simple.  

The complex zeros, $\zeta_0$ and $\zeta_p$, of $F(\zeta)$ and $F_{el}(\zeta)$, respectively, have been found numerically using Muller's algorithm \cite[p. 136]{DMYoung}. Note that the zeros will only depend on the cone angle $\theta_0$ via the weights $w_\ell(\theta_0)$ defined in (\ref{eq:check1})). Figure~\ref{fg:poles} shows an example distribution of the zeros $q_0 = j\zeta_0$ and $q_p=j\zeta_p$ in the complex $q = st_0=j\zeta$ plane for a cone angle $\theta_0 =65.22^o$. At this cone angle, the characteristic impedance of the infinite conical transmission line is $Z_0 = 53.6\,\Omega$. The conjugate pair nature of the zeros is clearly visible in Fig.~\ref{fg:poles}. It is also interesting to observe that the zeros of $F(\zeta)$ and $F_{el}(\zeta)$ are interlaced with respect to the imaginary $q$-axis (the real $\zeta$-axis). This is expected because those zeros are complex frequencies where the feed-gap reflection coefficient corresponds to a short circuit or an open circuit, respectively. It may be recalled that the frequencies at which a fixed length, terminated transmission line can resonate will alternate between a short circuited and an open circuited load.

It is possible to predict the number of zeros that exist within a semicircle of some fixed radius $q_R$ in the left half of the complex $q$-plane by resorting to the argument principle \cite[p. A-4]{RJ2020}. Equipped with these tools, we will now derive the time-domain responses for various circuit and field quantities. 
\begin{figure}[htb]
\centerline{\scalebox{0.45}{\includegraphics{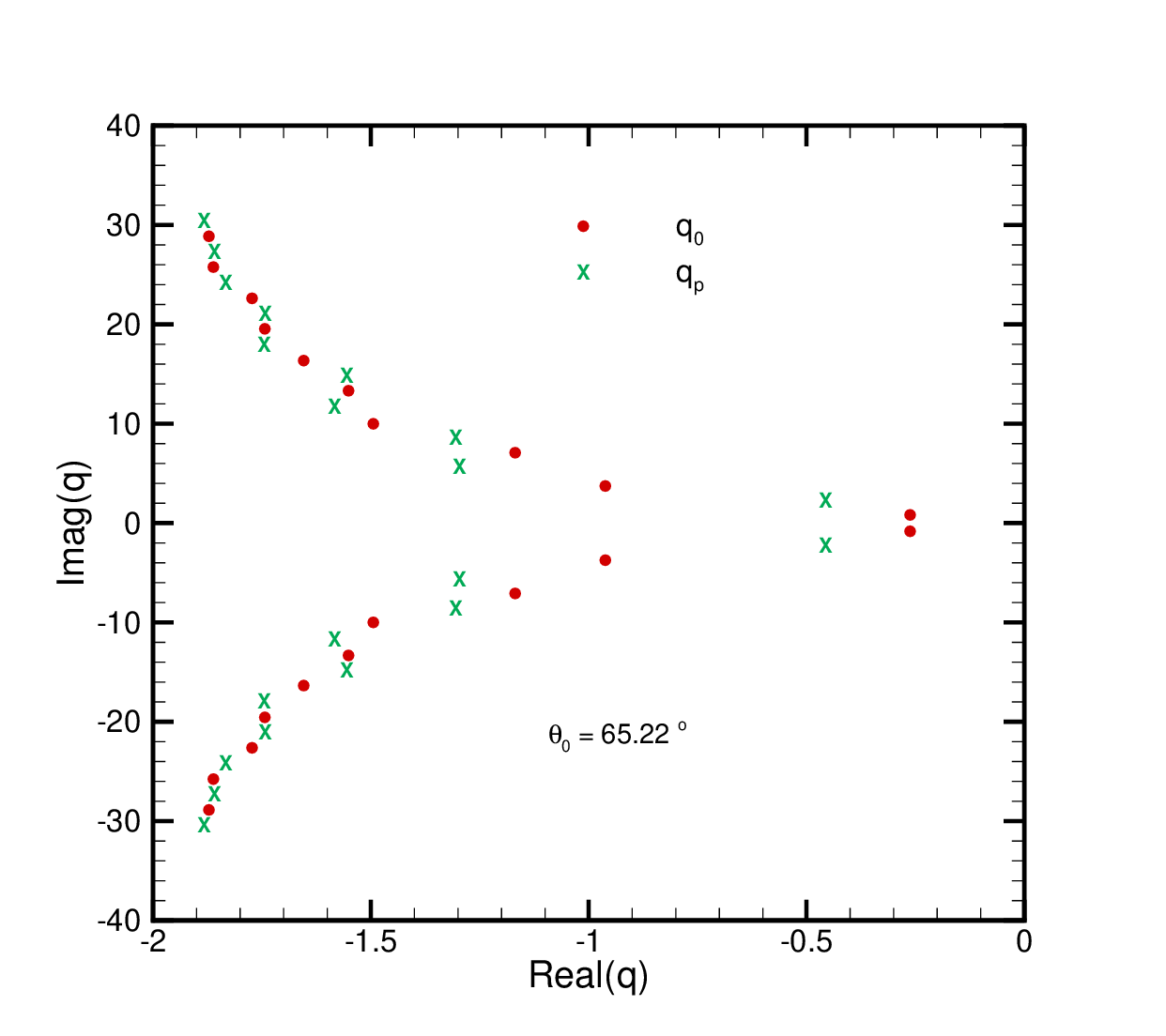}}}
\begin{center}
\caption{Distribution of zeros of $F(\zeta)$ and $F_{el}(\zeta)$ in the complex $q=j\zeta$ plane for a half-angle of $\theta_0 = 65.22^o$. They are interlaced with respect to the imaginary $q$-axis.}
\label{fg:poles}
\end{center}
\end{figure} 
\subsubsection{Impulse Response of Feed Current}
We are interested in the impulse response $i_\delta(\tau)$ for $\tau\gtrless \epsilon$. For $\tau<\epsilon$, the contour $\Gamma_q$ may be closed by a semi-circular arc $S_\infty^-$.  The integral yields a zero value because the integrand is analytic in the RHP. 
For $\tau>\epsilon$, we write 
\begin{equation}
\frac{F_{el}(\zeta)}{F(\zeta)} = \left[\frac{F_{el}(\zeta)}{F(\zeta)}-j\right]+j\label{eq:FelbyF}
\end{equation}
and close the contour $\Gamma_q$ by the semi-circular arc $S_\infty^+$ in the left-half-plane. The first part shown within the square brackets vanishes on $S_\infty^+$ by virtue of (\ref{eq:FeloverFinf}). Laplace inversion of the first part in (\ref{eq:FelbyF}) can be evaluated analytically by capturing residues at the poles of the integrand ({\em viz.\/}, at the zeros $\zeta_0$ of $F(\zeta)$). The residue of $F^{-1}(\zeta)$ at its pole $q=q_0\ (=j\zeta_0)$ is $j\zeta_0^2e^{q_0}/G(\zeta_0)$. The second part of (\ref{eq:FelbyF}) gives a value zero for $\tau>\epsilon$. Therefore for $\tau\lessgtr \epsilon$ we arrive at the result
\begin{equation}
t_0i_\delta(\tau) = V_0Y_0\sum\limits_{\zeta_0}\frac{\zeta_0^2\,e^{q_0(\tau-\epsilon+2)}}{G(\zeta_0)}\Theta(\tau-\epsilon),\label{eq:IRcur2}
\end{equation}
where $\Theta(\tau)$ is a unit step function. The residue is also seen to be independent of the antenna length $L$. In the rest of the paper, we will take $\epsilon\to 0$ without any loss of generality.

The impulse response is seen to be comprised of a sum of complex exponentials in the normalized temporal variable $\tau$. Furthermore, it is causal as it should be. The length of the antenna comes into picture through the transit time $t_0=L/c$ appearing in $\tau = t/t_0$. The real part of $q_0/t_0$ will dictate the temporal decay rate of the exponential in $t$, while its imaginary part will govern the oscillation rate in $t$.  A given root $q_0$ will then experience a smaller decay constant as well as a smaller oscillation frequency for a longer antenna than for a shorter one. Therefore more roots will manifest in early time in a longer antenna than in a shorter one. 

Since the dc feed current $I(0;\zeta=0) = 0$ by virtue of (\ref{eq:Gamlmts}), the integral of $i_\delta(\tau)$ over $\tau\in(\epsilon,\infty)$ must yield zero. Thus we have the identity
\begin{equation}
\sum\limits_{\zeta_0}\frac{\zeta_0\,e^{j2\zeta_0}}{G(\zeta_0)} = 0,\label{eq:check2}
\end{equation} 
which could serve as a guideline for determining the number of zeros required to achieve a given accuracy. 

Since the impulse response is a linear function of the source voltage amplitude $V_0$, we can use the principle of superposition to determine the response to any other input voltage $v_g(t)$ (units of [Vs$^{-1}$]):
\begin{equation}
i_g(\tau) = \int\limits_{-\infty}^\infty \frac{1}{V_0}t_0i_\delta(\tau-\tau')v_g(\tau')\,d\tau'\label{eq:isubg}\quad\mbox{[As$^{-1}$]}.
\end{equation}
The convolutional integral in (\ref{eq:isubg}) can be evaluated analytically in a number of situations owing to the presence of complex exponentials in $i_\delta(\tau)$. 
\subsubsection{Impulse Response of Radiated Field}
We write $kr = kL\, r/L = \zeta R=-jqR$ in (\ref{eq:rErad}), $R = r/L$, and consider the inverse Laplace transform of $rE_\theta$:
\begin{equation}
t_0r\mathcal{E}_\theta^\delta(r,\theta;\tau) = -\frac{V_0C_0}{2\pi j}
\int\limits_{\Gamma_q}\frac{F_e(\zeta)e^{q(\tau-R)}}{F(\zeta)}\,dq.
\end{equation} 
The units of ${\mathcal E}_\theta^\delta$ are [Vm$^{-1}$s$^{-1}$]. For $\tau-R<0$ the contour may be closed by $S_\infty^-$ to yield a zero value.  For $\tau-R>0$ the contour is closed on the left by $S_\infty^+$. Using the representation (\ref{eq:Fz0}) of $F(\zeta)$ near its zero and evaluating the integral by capturing the residues we arrive at the causal impulse response of the electric field as 
\begin{equation}
t_0r\mathcal{E}_\theta^\delta(r,\theta;\tau)=\Theta(\tau-R)V_0C_0\sum\limits_{\zeta_0}Q_e({\bf r};\zeta_0)e^{q_0(\tau-R+1)}\label{eq:IRErad}
\end{equation}
where
\begin{equation}
Q_{\left\{e\atop h\right\}}({\bf r};\zeta)=-\frac{j\zeta^2}{G(\zeta)}F_{\left\{e\atop h\right\}}({\bf r};\zeta).\label{eq:IRQ}
\end{equation}
The expression for the impulse response of magnetic field, $t_0r{\mathcal H}_\phi^\delta$, is obtained from (\ref{eq:IRErad}) by carrying out the replacements $V_0\to V_0/\eta,$ and $\ Q_e\to Q_h$. 

As with the feed current, the radiated field for any other input voltage $v_g(t)$ [Vs$^{-1}$] is given by the superposition integral 
\begin{equation}
r\mathcal{E}_\theta^g(r,\theta;\tau) = \int\limits_{-\infty}^\infty\frac{1}{V_0}t_0r\mathcal {E}_\theta^\delta(r,\theta;\tau-\tau')v_g(\tau')\,d\tau'.\label{eq:rEg}
\left[\frac{\rm V}{\rm s}\right]\end{equation}
\subsubsection{Temporal Receive Effective Length}
The temporal effective length of the antenna is obtained by taking the Laplace inverse of (\ref{eq:heff}). As already remarked, the relevant poles of the integrand here are the simple zeroes $\zeta_p$ ($=-jq_p$) of $F_{el}(\zeta)$. Evaluating the Laplace inverse by capturing the residues, the effective length is obtained as
\begin{equation}
t_0h_e^{Rx}(r,\theta;\tau) = -2L\sum\limits_{\zeta_p}\frac{Q_e({\bf r};\zeta_p)}{q_p}e^{q_p(\tau+1)}.\quad\mbox{[m]}\label{eq:hefft}
\end{equation}
Note that the temporal effective length of the antenna need not be positive for all times. 

It is seen that the formulas for the three temporal quantities described above are all analytical and can be evaluated once the expansions coefficients $c_\ell$ and the zeros $\zeta_0$ and $\zeta_p$ are determined numerically. And these parameters depend only on the half apex-angle $\theta_0$ of the antenna. Everything else in these formulas is  in terms of known special functions, which depend on frequency, range and aspect angle of the observation point.  

In terms of the effective length, the radiated electric field of the antenna for a general input can also be expressed using (\ref{eq:vecefflen}) as 
\begin{equation}
4\pi r{\mathcal E}_{\theta}^g({\bf r};\tau) =\eta\frac{d}{d\tau^\prime}\left[h_e^{Rx}({\bf r};\tau^\prime)\circledast i_g(\tau^\prime)\right]_{\tau'=\tau-R},\label{eq:Erad1}
\end{equation}
where $\circledast$ denotes convolution. For an ideal antenna having an impulsive effective length $h_e^{Rx}({\bf r};\tau^\prime) = h_0({\bf r})\delta(\tau^\prime)$, the radiated electric field becomes
\begin{equation}
4\pi r{\mathcal E}_\theta^g({\bf r};\tau) = \eta h_0({\bf r})\left.\frac{di_g(\tau^\prime)}{d\tau^\prime}\right\vert_{\tau^\prime=\tau-R}.\label{eq:Erad2}
\end{equation}
The radiated electric field is then the time-derivative of the feed current. {This time-derivative property is true only when the time-domain vector effective length of the antenna in its receive mode is proportional to an impulse function. Equivalently, the frequency domain vector effective length of the antenna must have all-pass characteristics (flat magnitude response and linear phase response). In \cite[p. 204]{Sengupta} an idealized, reflectionless dipole antenna (namely, one that supports a traveling wave current distribution) is shown to exhibit an approximate time-derivative property. For a non-ideal, but relatively broadband antenna, the radiated field will be approximately the time derivative of the feed current provided that the spectral content of its receive vector effective length exhibits a near ideal bandpass characteristics within the antenna bandwidth.} In practice, the non-ideality could be caused by a non-linear phase or a non-constant magnitude of the effective length over the spectral band of the input feed voltage pulse.   

There is only scant amount of well-documented experimental time-domain data available on the biconical antenna. Noteworthy exceptions in this regard are references \cite{LiculPhD} and \cite{Licul}, where the author conducts detailed measurements on several UWB antennas. Licul measured the frequency-domain antenna transmission coefficient by employing similar antenna elements at the transmitting and receiving ends and separated by a distance $r$. He converted the data to time-domain using inverse FFT and extracted the temporal effective length from it. Finally the temporal data for effective length was fit in terms of complex exponentials using the Prony's method \cite[p. 458]{Hildebrand}. In Fig.~\ref{fg:CompExp} we compare the theoretical effective length predicted by (\ref{eq:hefft}) with the measured results of Licul for a biconical antenna with $\theta_0 = 65.22^o,\ L =  17.2\,$cm, $r = 1.77\,m$. There was a small time offset of $t_{\rm off} = 31\,$ps\footnote{Results of \cite{LiculPhD} lagged our results by 31\,ps.} between the theoretical and measured result peaks, which we adjusted in Fig.~\ref{fg:CompExp}. In order for the measurement range of $r=1.77\,$m to correspond to far-zone distance for the antenna, its maximum linear dimension of $D = 2\times 17.2\,$cm and the smallest wavelength $\lambda_s$ in the spectrum must together satisfy $\lambda_s\ge 2D^2/r = 13.37\,$cm. In other words, the highest frequency content in the spectrum must be no more than 2.24\,GHz. However, Licul hinted at a baseband spectral bandwidth of 6\,GHz in the measurements. So the measurements he conducted are deemed to be in the near-zone. In this sense, the range dependent effective length we show in (\ref{eq:hefft}) becomes very relevant. The theoretical model used 26 complex zeros $\zeta_p$ and 101 expansion coefficients $c_\ell$. A very good agreement between the two is seen for the main pulse in Fig.~\ref{fg:CompExp}. It is also seen that the effective length takes negative values for some time durations. The deviations between the two results outside the main pulse could be attributed to (i) measurement errors that could possibly have resulted from noise when converting from frequency domain to time domain \cite{sarrazin}, and (ii) differences in the endcap geometry of the antenna used in reference \cite{LiculPhD} and here.\footnote{In \cite{LiculPhD} the biconical antenna was uncapped and hollow as opposed to the closed antenna with PEC spherical end-caps employed here.}
\begin{figure}[htb]
\centerline{\scalebox{0.45}{\includegraphics{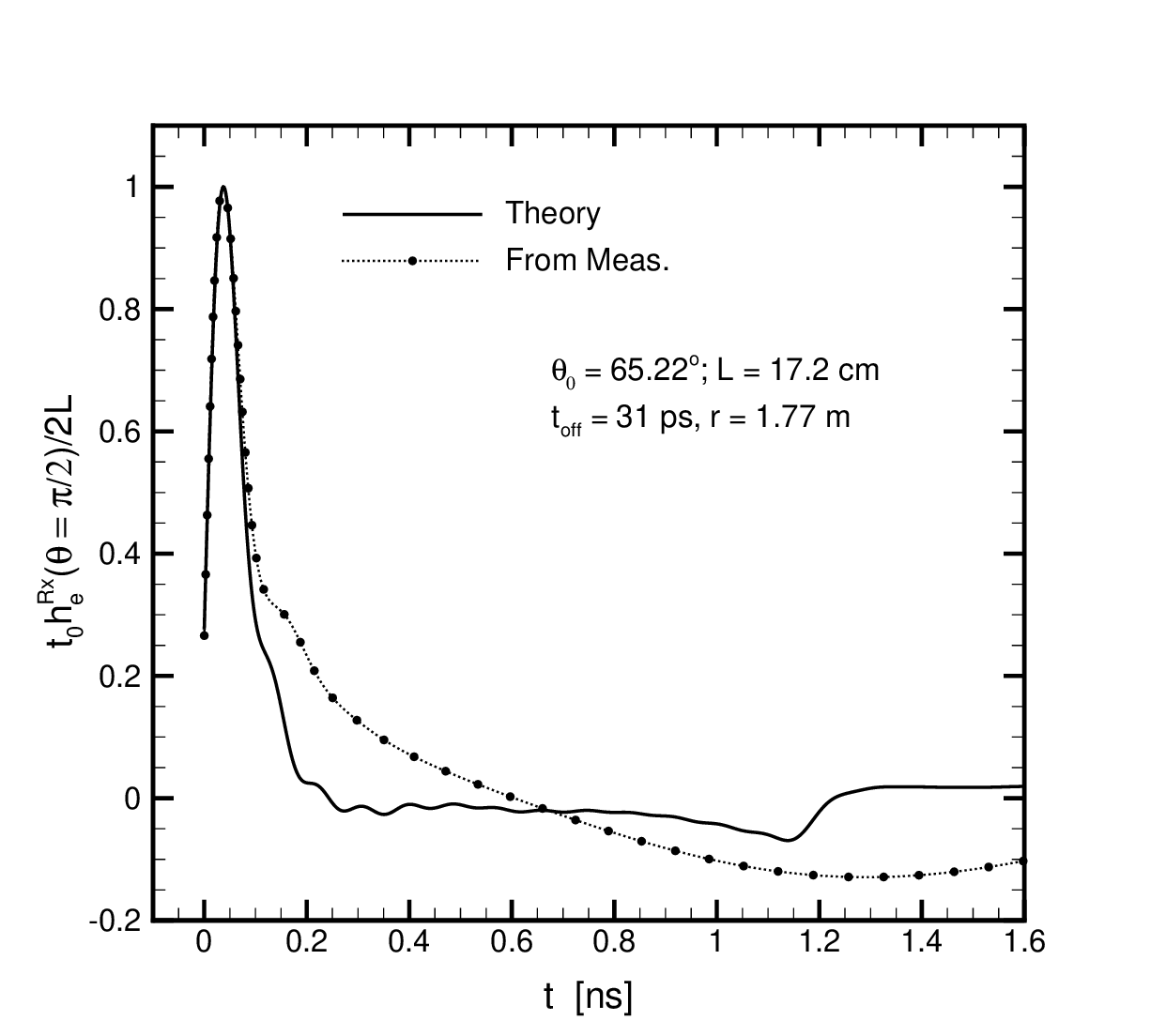}}}
\begin{center}
\caption{Comparison of temporal effective length with measured results of Licul \cite{LiculPhD}.}
\label{fg:CompExp}
\end{center}
\end{figure} 

Figure~\ref{fg:heofkL} shows a plot of the frequency-domain effective length in the far-zone and at boresight (that is, for $\theta = \pi/2$,  $r\to\infty$) for the antenna with same parameters as in Figure~\ref{fg:CompExp}. The magnitude has a dynamic range of about 16\,dB and the phase varies nonlinearly by about 1.4\,rad over $kL\in(0,20)$. The nonlinear phase variation coupled with the magnitude variation of the effective length does not render the antenna extremely wideband at this apex angle. However, beyond the knee occurring at around $kL = 4$, the effective length is relatively flat and the phase is approximately linear, barring some oscillations.  
\begin{figure}[htb]
\centerline{\scalebox{0.45}{\includegraphics{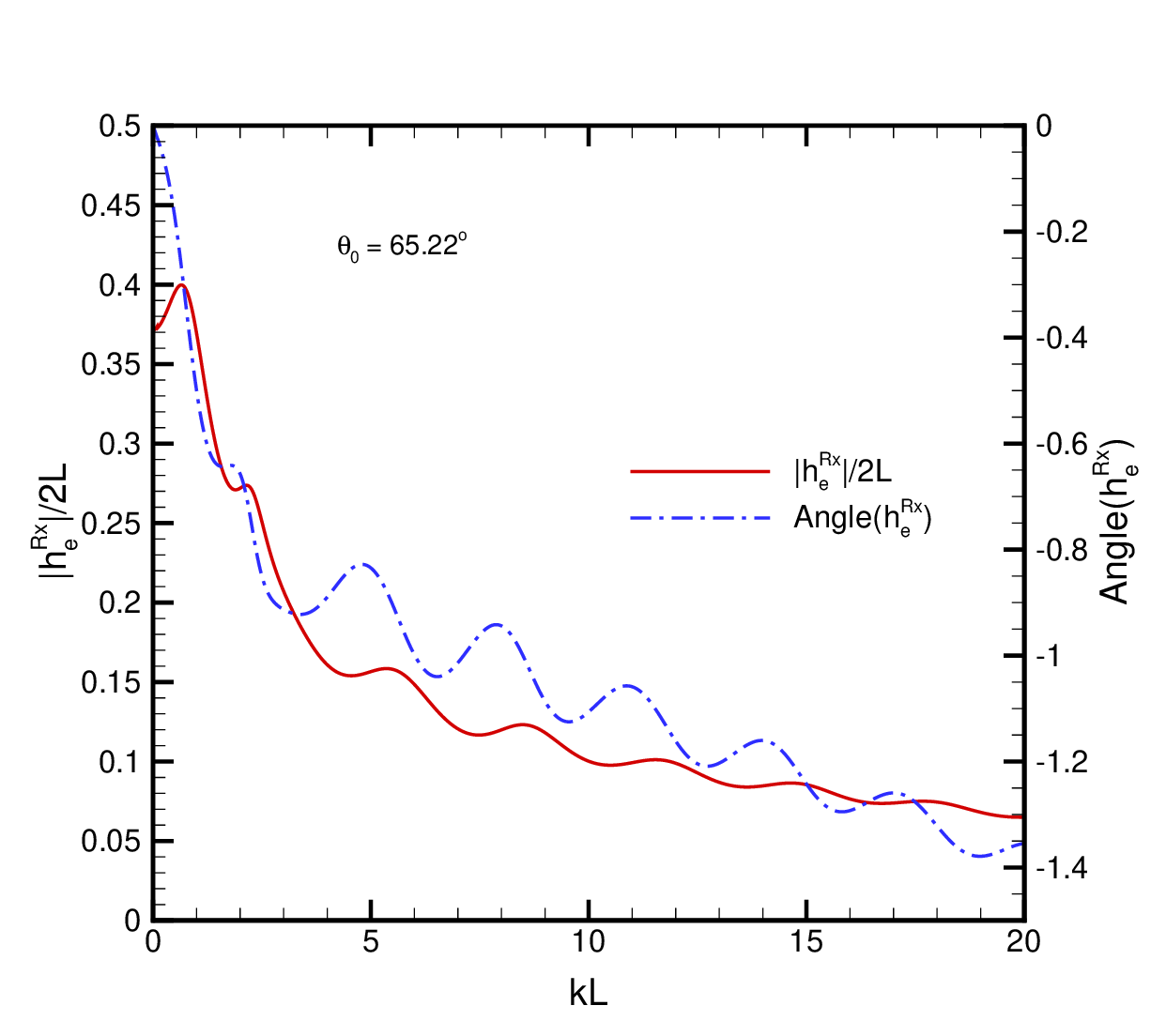}}}
\begin{center}
\caption{Magnitude and phase of antenna effective length, $h_e^{Rx}$, in the far-zone at boresight. Antenna length $L = 17.2\,$cm.}
\label{fg:heofkL}
\end{center}
\end{figure} 
\subsection{Response to Gaussian Input Voltage}
We consider a truncated (that is, one that spans over the semi-infinite interval $t>0$) Gaussian input voltage pulse of the form
\begin{equation}
v_g(t) = \frac{V_0}{\sqrt{\pi}\sigma_t}e^{-(t-t_c)^2/\sigma_t^2}\,\Theta(t),\quad\mbox{[Vs}^{-1}]\label{eq:Gauss1}
\end{equation}
where $\sigma_t\,$[s] characterizes the rise time as well as the pulse width. The Gaussian pulse assumes a maximum value at $t=t_c$ and a non-zero value at $t=0$. The Fourier transform of the pulse can be evaluated in a closed form and given in the normalized frequency $\zeta=kL$ as
\begin{equation}
V_g(\zeta) = 0.5V_0\,e^{-j\zeta\tau_c}\,e^{-\zeta^2\sigma^2/4}\,\erfc{\xi},\quad\mbox{[V]}
\end{equation}
where $\sigma=\sigma_t/t_0,\ \tau_c=t_c/t_0,\ \xi = j\zeta\sigma/2 -\tau_c/\sigma$ and $\erfc{\cdot}$ is the complementary error function \cite[7.1.2]{Abramo}.  Note that $v_g(0^+) >0$ and the dc component $V_g(0) = 0.5V_0\,\erfc{-\tau_c/\sigma}$. For $t_c/\sigma_t=\tau_c/\sigma>2$, $\abs{\erfc{\xi}}\approx 2$ and $\abs{V_g(\zeta)} \approx V_0\,e^{-\zeta^2\sigma^2/4}$. In that case the normalized 10\,dB bandwidth of the pulse is $\zeta_{\rm max}\approx 2.15/\sigma\approx $ (rise-time)$^{-1}$. The pulse width, $t_w$, can be taken to be $t_w\approx 2\sigma_t=2t_0\sigma$. 

We substitute (\ref{eq:Gauss1}) into (\ref{eq:isubg}) and (\ref{eq:rEg}) and evaluate the integrals in terms of the complex-valued complementary error function. The results are (recalling that $R = r/L$, $\tau = t/t_0$, $t_0 = L/c$, $q_0=j\zeta_0$)
\begin{eqnarray}
t_0i_g(\tau) &=&\Theta(\tau)\frac{V_0Y_0}{2}\sum\limits_{\zeta_0}\frac{q_0^2\,e^{q_0(\tau-\tau_c+2)}e^{(q_0\sigma/2)^2}}{G(\zeta_0)}\nonumber\\
&&\times\left[\erfc{\xi_1}-\erfc{\xi_0}\right],\label{eq:iGauss}\\
t_0r\mathcal{E}_\theta^g({\bf r};\tau)&=&\Theta(\tau-R)\frac{V_0C_0}{2}\sum\limits_{\zeta_0}Q_e({\bf r};\zeta_0)e^{q_0(\tau-\tau_c+1-R)}\nonumber\\
&&\times e^{(q_0\sigma/2)^2}\left[\erfc{\xi_2}-\erfc{\xi_0}\right],\label{eq:rEgGauss}
\end{eqnarray}
where 
\begin{equation}
\xi_0 = \frac{q_0\sigma}{2} -\frac{\tau_c}{\sigma};\ \xi_1 = \frac{\tau}{\sigma}+\xi_0;\ \xi_2 = \xi_1-\frac{R}{\sigma}.\label{eq:xis}
\end{equation}
\begin{figure}[htb]
\centerline{\scalebox{0.45}{\includegraphics{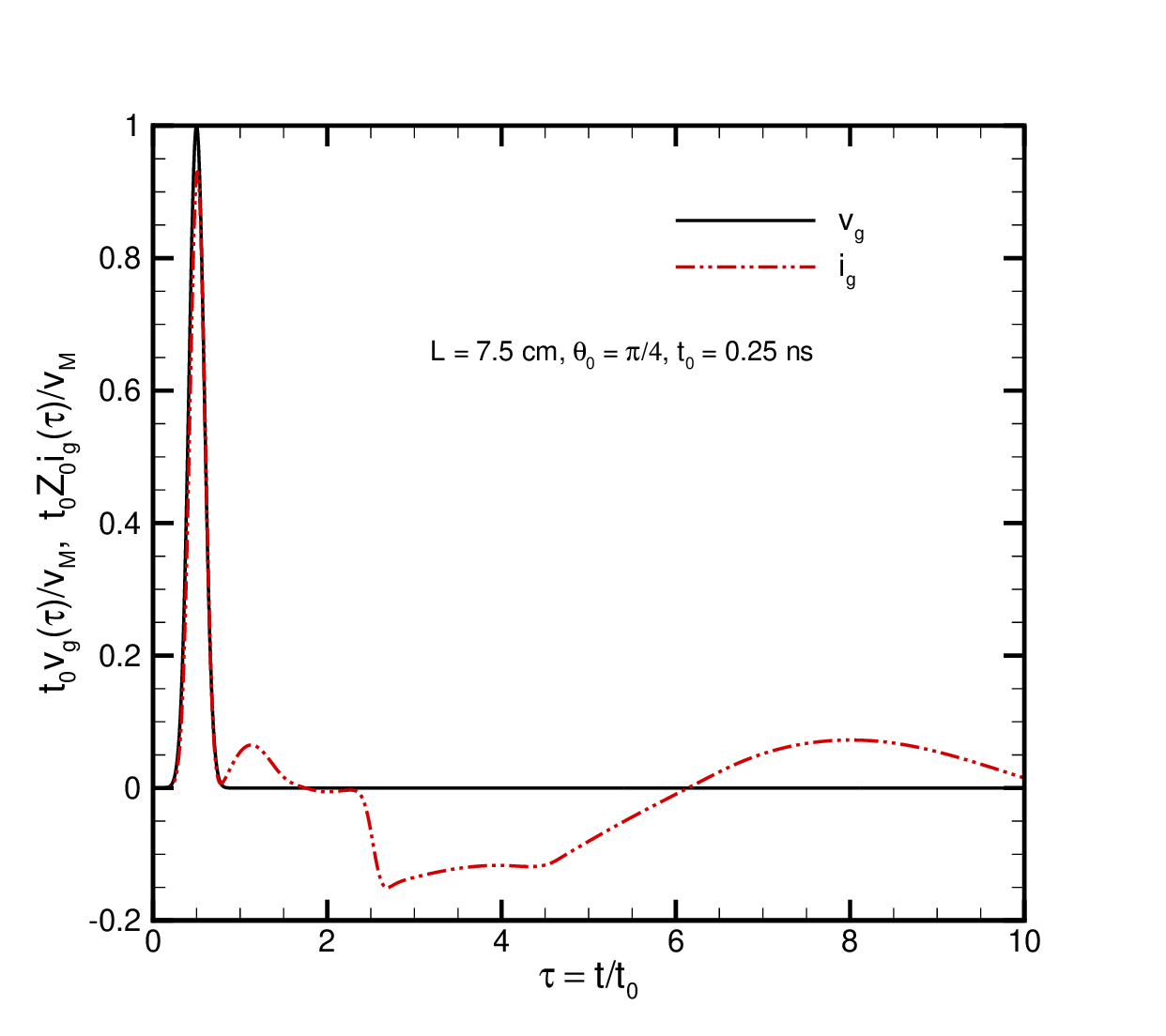}}}
\begin{center}
\caption{Input Gaussian voltage $v_g(\tau)$ (solid, black) and the resulting feed current $i_g(\tau)$ (dot-dashed, red).}
\label{fg:igoft}
\end{center}
\end{figure} 
Figure~\ref{fg:igoft} shows a plot of the transient feed current together with the input Gaussian voltage $v_g(\tau)$ for $\theta_0 = \pi/4$, $L = 7.5\,$cm, $\tau_c = 0.5$ and $\sigma = 1/8$.\footnote{In order to avoid inter-symbol interference Harrison \cite{Harrison} recommends that the significant pulse width must satisfy $t_w\le (1-\sin\theta_0)t_0$. In the present case $t_w = 2t_0\sigma$ and $\theta_0 = \pi/4$, requiring $\sigma\le 1/6.83$.}The transit time on the antenna length is $t_0 = L/c = 0.25\,$ns and the input pulse width is $t_w = t_0/4=62.5\,$ps. The normalization constant shown in the ordinate is $v_M = \max(t_0v_g(t))$. The initial pulse occurring in the feed current with a peak at $\tau=0.5$ replicates the input voltage pulse. A secondary negative peak is seen to arise at $\tau\approx 2.5$ ($\implies t \approx t_c + 2t_0$), which is due to reflection taking place at the far end of the antenna end and arriving back at the feed-gap. However its amplitude is only 1/6 of the amplitude of the primary peak, indicating that the reflections are small and that the feed current is primarily comprised of a forward propagating wave. The analytical results shown were generated using 101 expansion coefficients $c_\ell$ and 40 complex zeros $\zeta_0$.  
\begin{figure}[htb]
\centerline{\scalebox{0.45}{\includegraphics{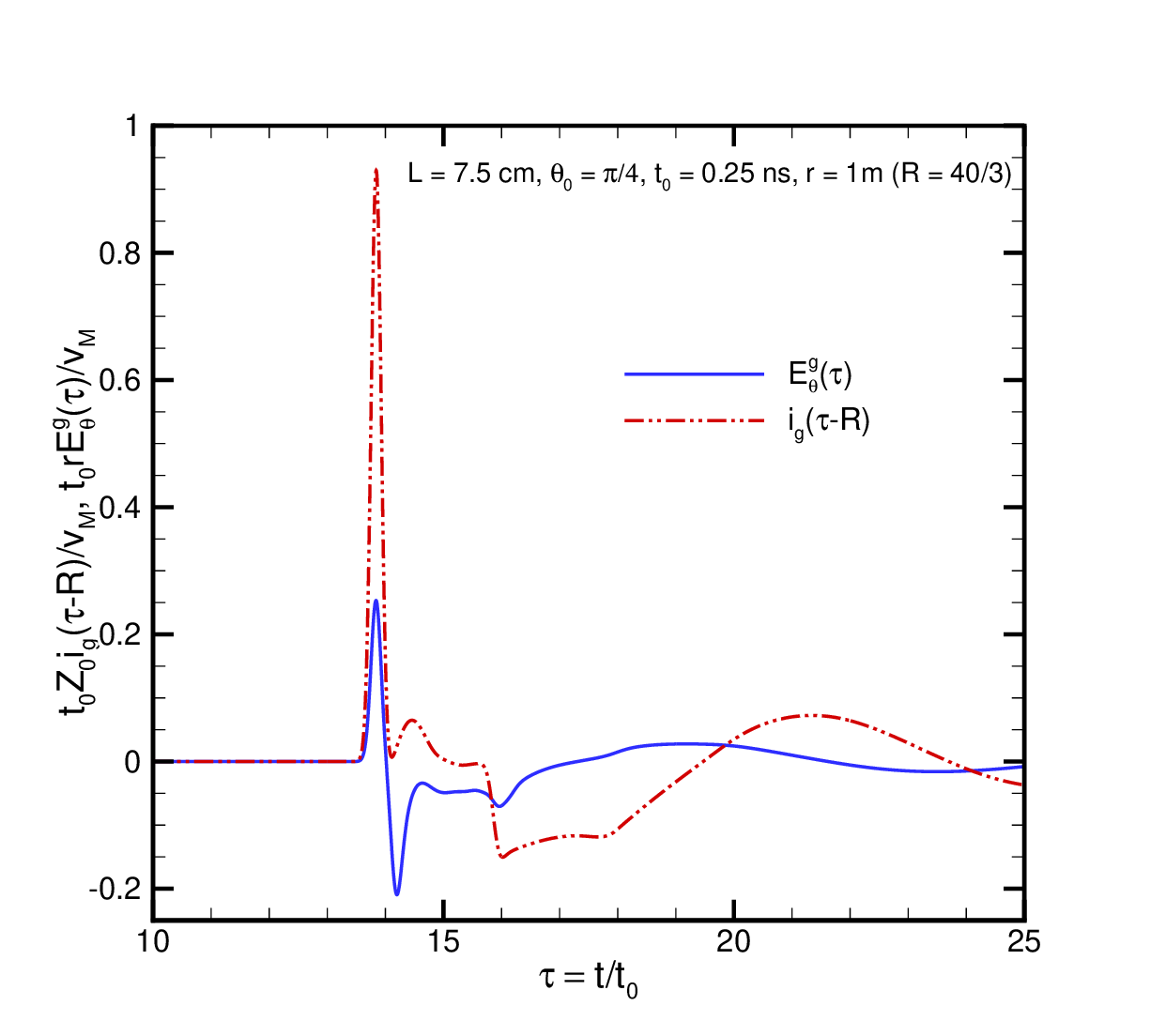}}}
\begin{center}
\caption{Radiated electric field $\mathcal{E}_\theta^g(\tau)$ (solid, blue) along boresight at fixed range $r$ and the delayed feed current $i_g(\tau-R)$ (dot-dashed, red) for comparison.}
\label{fg:rEthoft}
\end{center}
\end{figure} 

Figure~\ref{fg:rEthoft} shows a plot of the scaled radiated electric field $t_0r\mathcal{E}_\theta^g$ at a range of $r = 1\,$m along the boresight (that is, for $\theta = \pi/2)$ of the antenna with $\theta_0 = \pi/4$, $L = 7.5\,$cm, $\tau_c = 0.5$ and $\sigma = 1/8$. For comparison we also show the shifted feed current $i_g(\tau-R)$. All amplitudes have been normalized with the same constant $v_M$ as in Figure~\ref{fg:igoft} so that the relative amplitudes shown for scaled $i_g$ and $\mathcal{E}_\theta^g$ in the figure are true. The 10-dB bandwidth of the input Gaussian voltage pulse is $\zeta_{\rm max} \approx 2.15/\sigma \approx 17$ and the smallest significant wavelength contained in the pulse is $\lambda_s = 2\pi L/\zeta_{\rm max}= 2.74\,$cm. The far-field distance of the antenna at this wavelength is $r_{\rm ff} = 2(2L)^2/\lambda_s = 1.64\,$m $> r=1\,$m. The electric field shown in Figure~\ref{fg:rEthoft} is therefore in the transition region between near-zone and far-zone. {Fig.~\ref{fg:heff_Leq75mm} shows the magnitude and phase of the normalized effective length of the antenna over the significant spectral content $\zeta\in(0,\zeta_{\rm max})$ of the input voltage pulse. Over this wide spectral extent, the magnitude of the effective length exhibits a dynamic range of about 25\,dB and the phase is only approximately linear.} Still, the radiated field has the approximate shape of a doublet, which is the time-derivative of the {\em primary} Gaussian pulse present in the feed current. Hence for the geometric parameters chosen, the antenna is performing approximately as an ideal UWB antenna. 
\begin{figure}[htb]
\centerline{\scalebox{0.45}{\includegraphics{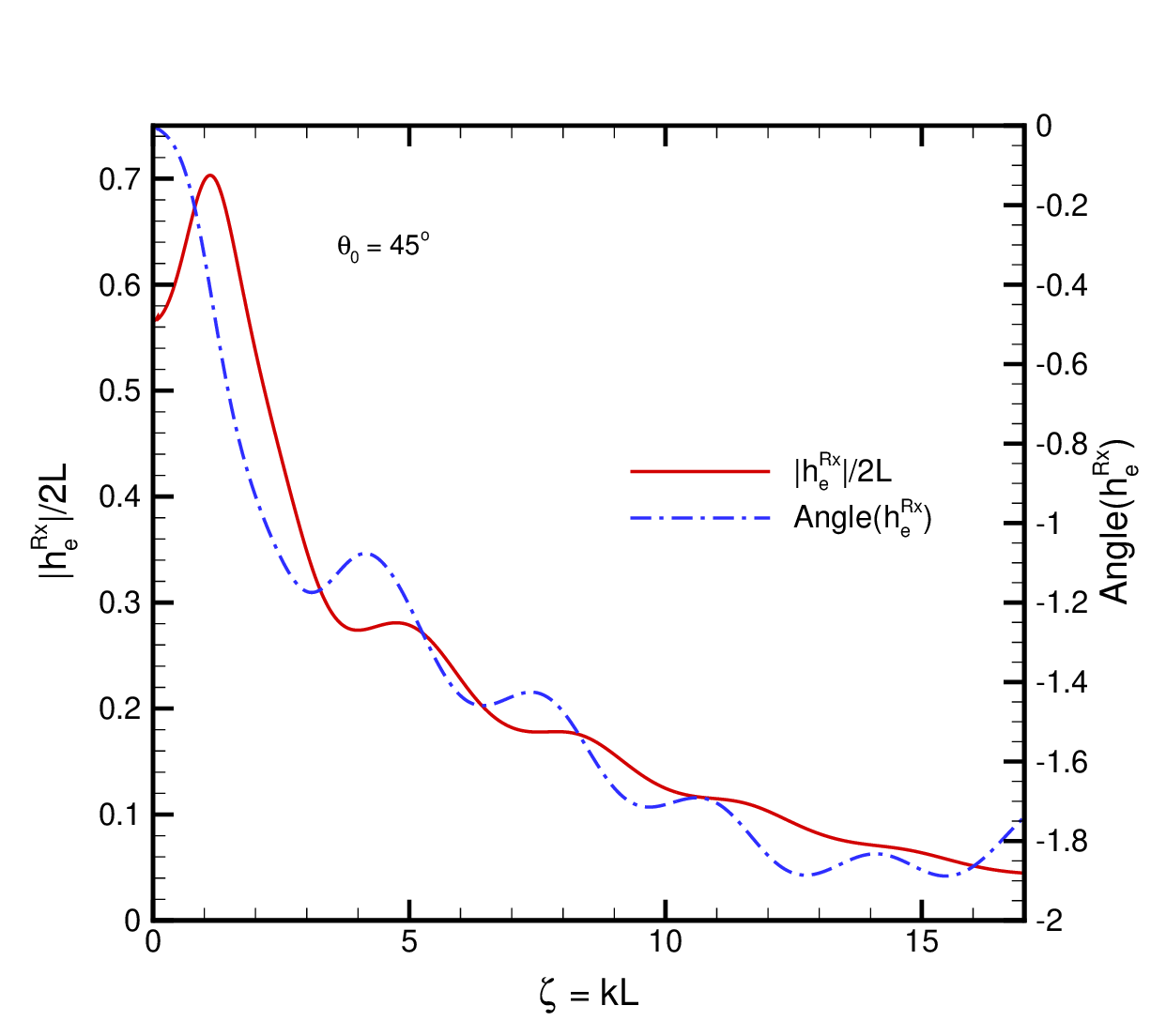}}}
\begin{center}
\caption{Magnitude and phase of the antenna effective length $h_e^{Rx}(\theta,r;\zeta)$ for $L = 7.5\,$cm, $\theta = \pi/2,\, r = 1\,$m.}
\label{fg:heff_Leq75mm}
\end{center}
\end{figure} 

\subsection{Radiated Energy and Missile-generating Voltage Pulse} 
The Gaussian input voltage pulse is not necessarily the ideal waveform that can be used with a biconical antenna because it decays too rapidly at high frequencies and, consequently, need not generate the narrowest possible radiated pulse for a given input energy. Here we explore a different waveform, while still not ideal, that will be better suited to the high frequency behavior of the antenna and provides some interesting radiative properties. For a non-dissipative antenna, the input energy provided at the antenna input terminals equals the {\em total\/} energy radiated, which in terms of the fields is 
\begin{eqnarray}
W_r &=& \int\limits_{\Omega_s}d\Omega\int\limits_{-\infty}^\infty t_0r\mathcal{E}_\theta({\bf r};t) t_0r\mathcal{H}_\phi({\bf r}:t)\,dt\nonumber\\
&=& \frac{t_0}{2\pi}\Re\left[\int\limits_{\Omega_s}d\Omega\int\limits_{-\infty}^\infty rE_\theta({\bf r};\zeta)rH_\phi^*({\bf r};\zeta)\,d\zeta\right],\label{eq:Enrgy1}
\end{eqnarray}
where $\Re(\cdot)$ denotes real part and $\Omega_s$ is the unit sphere in 3D. The latter equality above follows from Parseval's relation. Substituting (\ref{eq:rErad})-(\ref{eq:Fz}) and (\ref{eq:rEg}) with a frequency domain generator voltage $V_g(\omega)$ into (\ref{eq:Enrgy1}) and carrying out the integral w.r.t. $\Omega_s$ we arrive at
\begin{eqnarray}
W_r &=&\frac{t_0}{\eta}\int\limits_{-\infty}^\infty V_g(\omega)\,T_J(\zeta)\,V_g^*(\omega)\,d\zeta,\label{eq:Enrgy2}\\
\noalign{where}
T_J(\zeta) &=&\frac{2C_0^2}{\abs{F(\zeta)}^2}\left[\sum\limits_{\ell=1,3}^\infty\frac{c_\ell^2\ell(\ell+1)/(2\ell+1)}{\abs{\widehat{H}_\ell^{(2)}(\zeta)+j\widehat{H}_\ell^{(2)\prime}(\zeta)}^2}\right]\nonumber\\
&=:&\frac{2C_0^2}{\abs{F(\zeta)}^2}\,G_J(\zeta),\label{eq:TJzet}
\end{eqnarray}
with $G_J(\zeta)$ corresponding to the term within the square brackets. The expression in (\ref{eq:TJzet}) is, as expected, independent of the range $r$. The energy transfer function, $T_J(\zeta)$, is positive semidefinite along the real frequency axis and assumes a value zero only at zero frequency. Figure~\ref{fg:TJzet} shows a plot of $T_J(\zeta)$ versus $\zeta = \omega t_0=kL$ for the antenna with a half-angle $\theta_0 = \pi/4$. Firstly, it is seen that the energy transfer function exhibits a {\em high-pass characteristic}. The peak at $\zeta =\zeta_r\approx  0.8$ is due to the first resonance of the biconical antenna \cite{janaswamy2022}. Beyond the first resonance, $T_J(\zeta)$ is seen to be relatively flat. Also notice that it rises as a power law, $\zeta^n$, at low frequencies with an exponent of $n = 4$, reminiscent of {\em Rayleigh scattering} \cite[p. 323]{Ishimaru}. This fourth power behavior at low-frequencies can also be demonstrated directly using (\ref{eq:Fzlmts}) and the small argument forms of the Hankel functions contained in $G_J(\zeta)$. At high frequencies $T_J(\zeta)\sim O(1)$. Therefore the total energy $W_r$ radiated by the antenna is not necessarily finite unless the input voltage $V_g(\omega)$ decays sufficiently at infinity.  For instance, with an impulse input voltage, the total energy radiated by the biconical antenna is infinite. 
\begin{figure}[htb]
\centerline{\scalebox{0.45}{\includegraphics{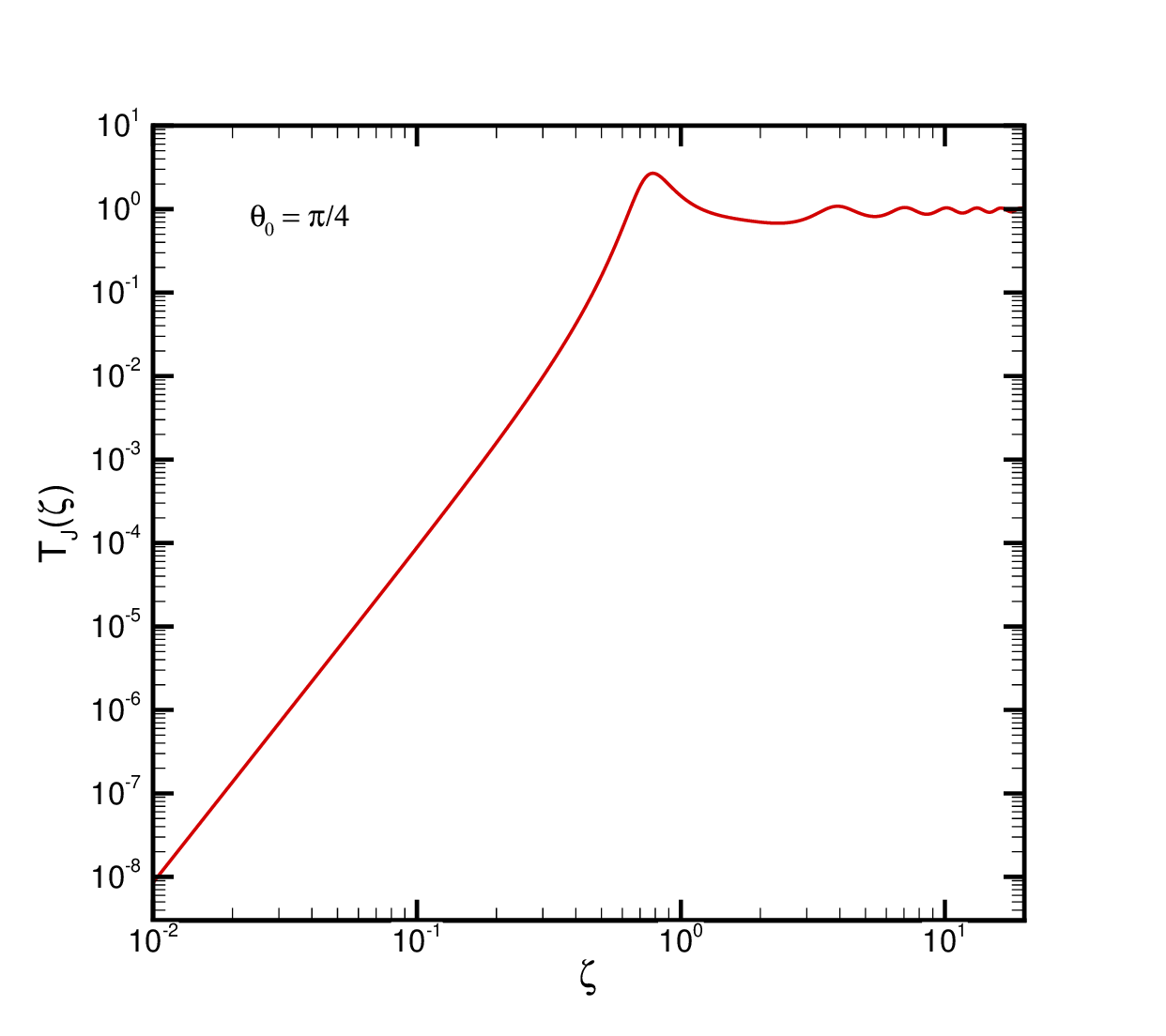}}}
\begin{center}
\caption{Energy transfer function $T_J(\zeta)$ of biconical antenna for $\theta_0 = \pi/4$.}
\label{fg:TJzet}
\end{center}
\end{figure} 

With the availability of analytical expressions for various quantities, it might be tempting to consider the optimization techniques outlined in \cite[Ch. 6]{Franks}, \cite[Ch. 4]{Pierre},\cite{Pozar} and design an input voltage waveform to achieve a specific goal. Unfortunately, these techniques are all geared towards {\em low-pass transfer characteristics\/} and do not apply to a biconical antenna. It is clear from (\ref{eq:Enrgy2}) that the input voltage $V_g(\omega)$ needs to have at least the algebraic decay behavior of $\abs{\omega}^{-(0.5 +\epsilon)}$, $\epsilon>0$ at infinity to maintain finite radiated energy. We consider the following waveform proposed in \cite{Hafizi} for generating an electromagnetic missile \cite{Hao-Ming} (that is, a radiated waveform that has a pathloss decay factor slower than $1/r^2$ in some region of space), while radiated from an ideal planar aperture\footnote{We modify the waveform of \cite{Hafizi} to fit our normalized time notation}:
\begin{equation}
v_g(\tau) =  V_0\omega_0\,\tau^{2\alpha}\,e^{-\tau/\tau_1}\,\frac{\sin(\Omega_0 \tau)}{\Omega_0\tau}\Theta(\tau),\quad\left[\frac{\mbox{V}}{\mbox{s}}\right]\label{eq:Haf1}
\end{equation}
where $\Omega_0 = \omega_0t_0$ and as usual $\tau = t/t_0$. The real-valued parameter $\alpha$ controls the rise-time behavior of the pulse near $\tau=0$ and the exponential time constant $\tau_1$ along with the sinc function factor govern the fall-time of the pulse. For $\alpha>0$, the initial value $v_g(0^+) = 0$. The Fourier transform of (\ref{eq:Haf1}) can be evaluated in a closed form\footnote{We express $V_g(\omega)$ in a form very different from that in \cite{Hafizi} to facilitate easier asymptotic and limiting forms.}
\begin{eqnarray}
V_g(\omega)&=& \frac{V_0}{2j}\frac{\Gamma(2\alpha+1)}{(s_1^2(\zeta)+\Omega_0^2)^\alpha}\frac{\sinh\left[\alpha\log\left(\frac{s_1(\zeta)+j\Omega_0}{s_1(\zeta)-j\Omega_0}\right)\right]}{\alpha}\ \ \ \label{eq:Haf2}\\
&\sim&\frac{V_0}{2j}\log\left(\frac{s_1(\zeta)+j\Omega_0}{s_1(\zeta)-j\Omega_0}\right),\quad \alpha\to 0\label{eq:Haf3}\\
&\sim&\frac{V_0\Omega_0\Gamma(2\alpha+1)}{(j\zeta)^{(2\alpha+1)}},\quad \abs{\zeta}\to\infty\label{eq:Haf4}\\
&\sim&O(1),\quad\zeta\to 0\label{eq:Haf5} 
\end{eqnarray}
where $s_1(\zeta)=\tau_1^{-1}+j\zeta$, $\zeta = \omega t_0=kL$ and $\Gamma(x)$ is the Gamma function. Expressions (\ref{eq:Haf1}) and (\ref{eq:Haf2}) are both valid for any $\alpha>-1/2$. However, to meet the finite radiated energy requirement we impose $\alpha>-1/4$. For $2\alpha$ not an integer, there will be branch points in the complex frequency plane at $q = -\tau_1^{-1}\pm j\omega_0$. 

We use (\ref{eq:isubg}), (\ref{eq:IRErad}) and (\ref{eq:rEg}) and determine the feed current and the radiated electric field in terms of the incomplete gamma function \cite[3.381-1]{GrRy2007} as
\begin{eqnarray}
t_0Z_0i_g(\tau)&=&\Theta(\tau)\frac{V_0\tau^{2\alpha}}{2}\sum\limits_{\zeta_0}\frac{jq_0^2}{G(\zeta_0)}e^{q_0(\tau+2)}\nonumber\\
&&\times\left[H_{2\alpha}(\beta_1(\tau))-H_{2\alpha}(\beta_2(\tau))\right]
\label{eq:Haf6}\\
t_0r{\mathcal E}_\theta({\bf r};\tau)&=&\frac{V_0C_0(\tau-R)^{2\alpha}}{2j}\sum\limits_{\zeta_0}Q_e({\bf r};\zeta_0)e^{q_0(\tau-R+1)}\nonumber\\
&&\times\left[H_{2\alpha}(\beta_1(\tau-R))-H_{2\alpha}(\beta_2(\tau-R))\right]\nonumber\\
&&\times\Theta(\tau-R),\label{eq:Haf7}
\end{eqnarray} 
where $\beta_{1\atop 2} = \tau_1^{-1}-j\Omega_0\mp q_0$ and 
\begin{equation}
H_\alpha(x) = \sum\limits_{n=0}^\infty\frac{(-1)^nx^n}{n!(n+\alpha)}\label{eq:Haf8}
\end{equation}
is related to the incomplete gamma function $\gamma(\alpha,x)$ via $H_\alpha(x) = x^{-\alpha}\gamma(\alpha,x)$ \cite[8.354]{GrRy2007}. Despite the presence of complex numbers on the RHS of (\ref{eq:Haf6}) and (\ref{eq:Haf7}), the time-instantaneous feed current and the radiated fields will be real-valued. The function $H_\alpha(x)/\Gamma(\alpha)$ is analytic w.r.t. $\alpha$ and $x$ \cite[8.351-1]{GrRy2007}. A similar expression is obtained for the magnetic field $t_0r{\mathcal H}_\phi({\bf r};\tau)$ after replacing the function $Q_e({\bf r};\zeta_0)$ in (\ref{eq:Haf5}) with $Q_h({\bf r};\zeta_0)$ and $V_0$ with $V_0/\eta$.

Figure~\ref{fg:HafCur} shows a plot of the voltage (\ref{eq:Haf1}) and the corresponding feed current (\ref{eq:Haf6}) as a function of time for an antenna with $L=75\,$cm, $\theta_0 = \pi/4$. The transit time $t_0 = 2.5\,$ns. The source parameters are $\alpha = 10^{-3}, \tau_1 = 0.5, \Omega_0 = 5\pi$. The number of modes $c_\ell$ and the number of zeros $\zeta_0,\,\zeta_p$ used in the analysis were 101, 40 and 26 respectively. Compared to the feed current of Figure~\ref{fg:igoft} induced by a Gaussian source, it is seen that the reflections here are somewhat smaller. But this could also be due to the fact that the antenna is ten times longer here. Before the onset of reflections at $\tau=2$, it is, however, seen that the feed current is a better reproduction of the source voltage here than with the Gaussian source voltage. Figure~\ref{fg:VgHaf} shows the normalized input source voltage $V_g(\omega)$ in the frequency domain. The peak of the voltage takes place at $\zeta = \Omega_0$ and the $-3\,$dB bandwidth is $\zeta_{\mbox{\textsubscript{3dB}}} \approx 19$. The far-field distance corresponding to this highest baseband frequency is $r = 4\,\zeta_{\mbox{\textsubscript{3dB}}}L/\pi \approx 24L$. 

\begin{figure}[htb]
\centerline{\scalebox{0.45}{\includegraphics{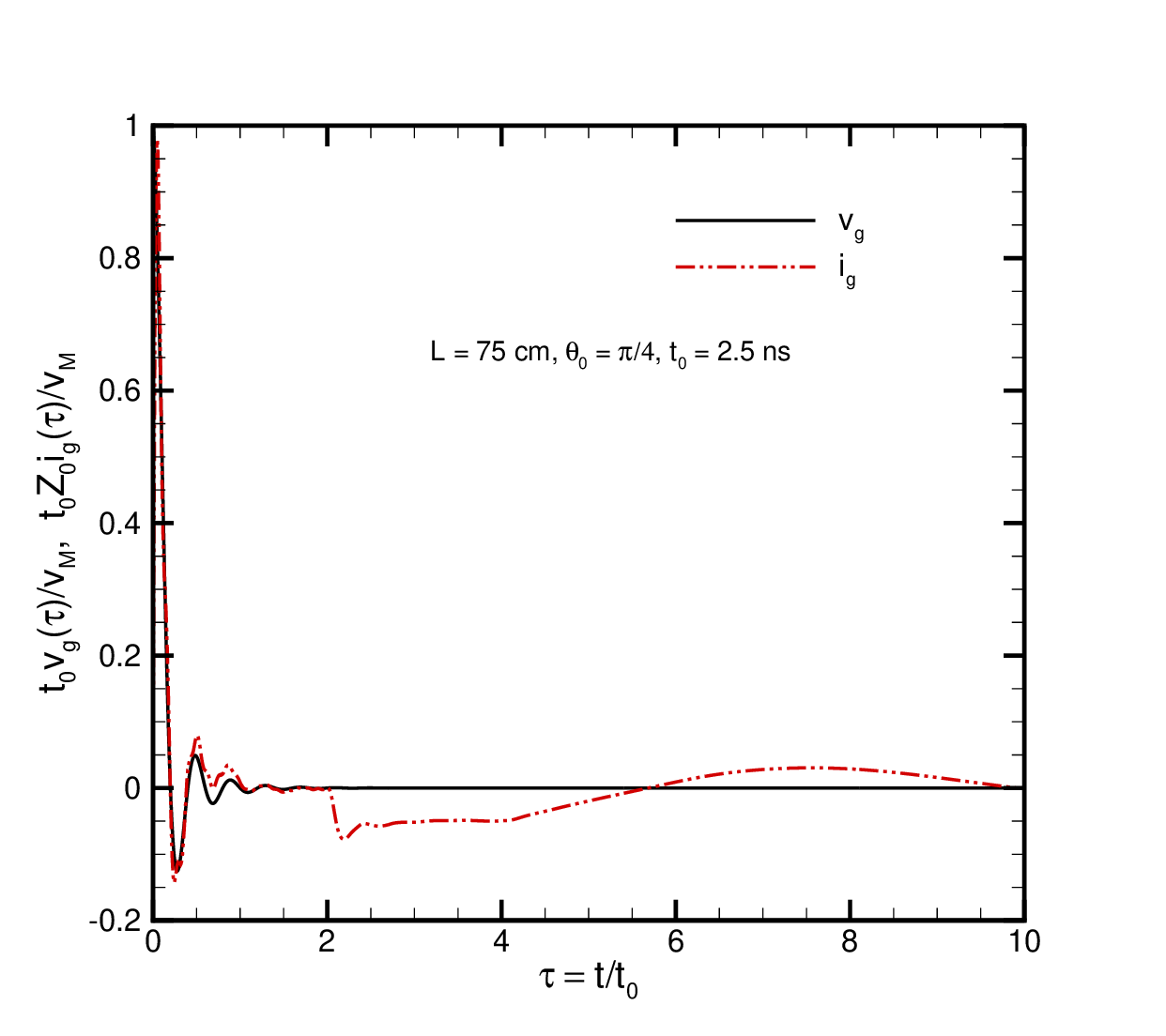}}}
\begin{center}
\caption{Input Gaussian voltage $v_g(\tau)$ (solid, black) and the resulting feed current $i_g(\tau)$ (dot-dashed, red) for the pulse given in (\ref{eq:Haf1}).}
\label{fg:HafCur}
\end{center}
\end{figure} 

In order to explore the potential of this waveform to generate an electromagnetic missile we look at the temporal energy distribution as a function of distance from the antenna in a particular direction. Fig.~\ref{fg:HPulse} shows a plot of the energy density $t_0^2r^2{\mathcal E}_\theta({\bf r};\tau){\mathcal H}_\phi({\bf r};\tau)\,$[Js$^{-1}$] along the boresight ${\bf r} = (r,\theta=\pi/2)$ as a function of $t = \tau t_0$ with $r$ as a parameter for an antenna with $L = 75\,$cm, $\theta_0 = \pi/4$. We arbitrarily take the constant $V_0 = \Omega_0^{-1}\left(\frac{e}{2\alpha\tau_1}\right)^{2\alpha}$. In the far-zone the energy density peak should remain constant with distance. The observation distance here is varied starting from $r=L$ and progressively increased in steps of $\Delta r = 2L = 1.5\,$m. At the 13th step the observation distance is $r = L + 12\times \Delta r = 25L$, which is at the edge of far-zone at the 3\,dB frequency.  In the immediate vicinity of the antenna ($L\le r\le 7L$)\footnote{The slight delay between the electric field and the magnetic field in the near-zone can cause the energy density to be negative at certain times. However, the time-integral of energy density is always positive as it should be.}, it is seen that the peak energy density increases with distance, implying that the radiated fields decay slower than the free-space decay of $1/r$ in the {\em near-zone}. Such slow decay of the near fields with practical antennas has also been demonstrated previously with reflector antennas fed by a unit-step spherical wave \cite{Chou}. 

\begin{figure}[htb]
\centerline{\scalebox{0.45}{\includegraphics{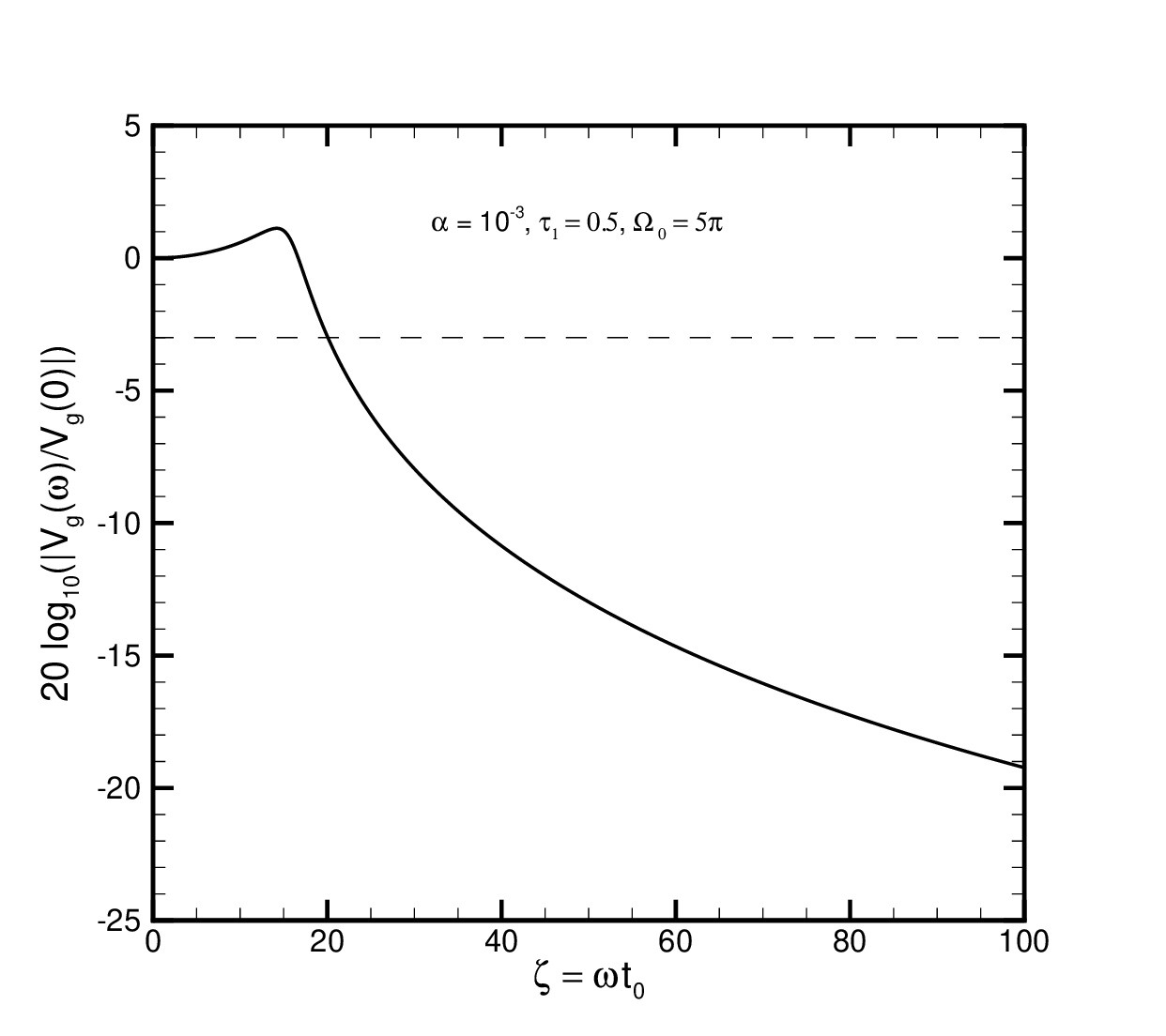}}}
\begin{center}
\caption{Normalized source voltage $20\log_{10}\abs{\frac{V_g(\omega)}{V_g(0)}}$ of (\ref{eq:Haf2}) versus $\zeta$.}
\label{fg:VgHaf}
\end{center}
\end{figure} 

\begin{figure}[htb]
\centerline{\scalebox{0.45}{\includegraphics{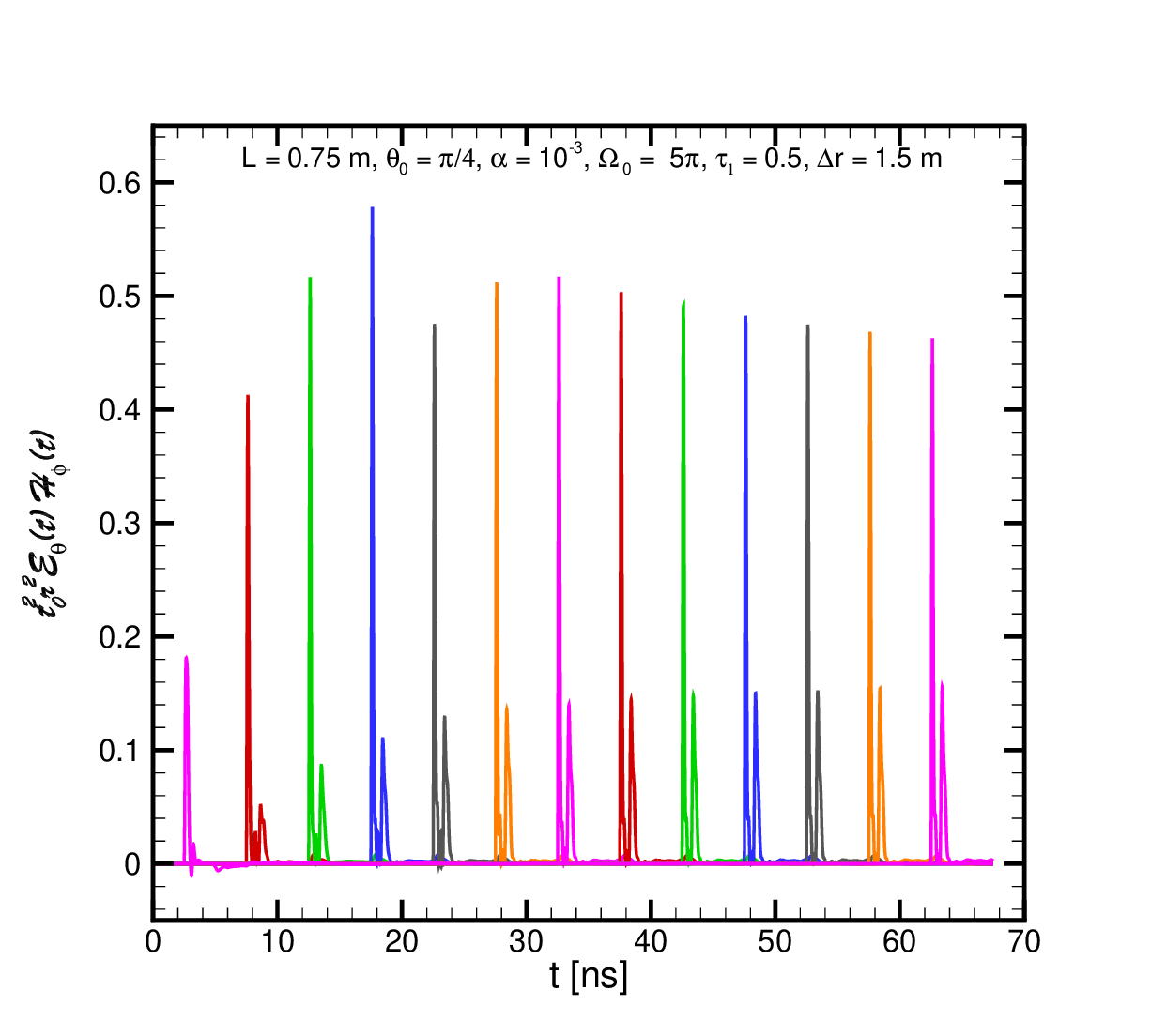}}}
\begin{center}
\caption{Variation of time-instantaneous radiated energy density $t_0^2r^2{\mathcal E}_\theta(r;t){\mathcal H}_\phi(r;t)$ of biconical antenna with distance along boresight. Spatial increment between pulses $\Delta r = 1.5\,$m.}
\label{fg:HPulse}
\end{center}
\end{figure}

\section{Summary and Conclusions}
Exact analytical expression have been derived for various antenna parameters in time-domain for the biconical antenna of arbitrary angle $\theta_0$ and length $L$. The biconical antenna has many desirable UWB features and the following items summarize the key steps/findings of this study:

\begin{enumerate}[(i)]
\item A new factorization of the spherical harmonic mode expansion coefficients as encompassed in (\ref{eq:A0}), (\ref{eq:rEr2})-(\ref{eq:rHph2}) and (\ref{eq:coeffcl}) is used in the paper to facilitate direct application of the Laplace inversion formula to arrive at analytical forms for various antenna properties.  The analysis presented in the paper is exact for any length and half-angle of the biconical antenna.  
\item The linear system (\ref{eq:cell2Eqn}) can be used to determine the frequency independent coefficients of spherical expansion. The coefficients are dependent only on the cone angle $\theta_0$. 
\item Complex zeroes $\zeta_0$ and $\zeta_p$ of the functions $F(\zeta)$ and $F_{el}(\zeta)$, respectively, can be found effectively using Muller's algorithm \cite[p. 136]{DMYoung}. The singularities of various functions are all shown to be {\em simple poles\/} in the left-half of the complex frequency plane ($s$-plane), confirming a conjecture made in the singularity expansion method \cite{CEBaum}. 
\item Once the expansion coefficients and the complex zeros are found, equations (\ref{eq:IRcur2}), (\ref{eq:IRErad}), and (\ref{eq:hefft}) can be used to find the impulse response of feed current, impulse response of radiated electric field, and the transient effective length, respectively. 
\item Prediction of transient effective length by the analytical formulas has been checked by favorable comparison with measurements. 
\item Equations (\ref{eq:isubg}) and (\ref{eq:rEg}) can be used to find the transient feed current and transient radiated electric field for an arbitrary input voltage pulse $v_g(t)$ in terms of their respective impulse responses. Analytical results have been provided for a Gaussian pulse (\ref{eq:Gauss1}) and a fast rising pulse of (\ref{eq:Haf1}). Radiated electric field in the far-zone behaving approximately as the time derivative of the feed current has been demonstrated in the response to the Gaussian voltage pulse. Biconical antenna fed by the fast rising pulse is shown to be capable of generating an electromagnetic missile in the near-zone. 
\item Finding feed voltage pulses necessary to optimize a given metric of radiation, while subject to practical constraints, will be the topic of future research. Some relevant time-domain functionals that could be useful in such studies are highlighted in \cite{SFWang}. 
\end{enumerate}


%
\appendices
\renewcommand*{\thesection}{\Alph{section}}
\setcounter{section}{0}
\section{Solution for Expansion Coefficients}
\label{sc:App1}
The coefficients $a_{\nu_n}(k)$ and $b_\ell(k)$ of field expansions can be determined by imposing the continuity conditions on the field $(E_\theta, H_\phi)$ at $r=L$ and testing with the relevant spherical harmonic function in a manner similar to what was done in \cite{janaswamy2022}. The identities \cite[(24)-(28)]{janaswamy2022} become useful in evaluating various integrals. Several observations are in order in this regard:
\begin{enumerate}[(i)]
\item The matching conditions are independent of the factor $A_0(k)$.
\item The coefficients $a_{\nu_n}(k), b_\ell(k)\to 0$ as $kL\to\infty$.
\item Angular functions satisfy $M_{\nu_n}(\theta_0) = 0 = M_{\nu_n}(\pi-\theta_0)$ and $P_\ell(\cos\theta_0) = -P_\ell(-\cos\theta_0)$. 
\item The composite field $E_\theta+\eta H_\phi$ will not explicitly involve $\Gamma_{\rm in}(k)$;  however, the field $E_\theta-\eta H_\phi$ will. These are also clear from the TEM field counterparts (\ref{eq:ETEM}) and (\ref{eq:HTEM}). 
\end{enumerate}
The coefficients $a_{\nu_n}(k)$ can then be expressed in terms of a series involving $b_\ell(k)$ and vice-versa. When these are combined, the following equation is obtained for the determination of the coefficients $b_m(k)$, $m = 1,3,\dots$
\begin{eqnarray}
j^{-m}\left[\widehat{H}^{(2)}_m(kL) + j\widehat{H}_m^{(2)\prime}(kL)\right]\frac{m(m+1)}{(2m+1)}b_m = \frac{2e^{-jkL}}{jkL}\times&&\nonumber\\ 
\times P_m(\cos\theta_0)+\sum\limits_{\ell=1,3,\ldots}^\infty j^{-\ell} \left[\widehat{H}^{(2)}_\ell(kL) + j\widehat{H}_\ell^{(2)\prime}(kL)\right]b_\ell \times&&\nonumber
\\
\times \sin\theta_0P_m(\cos\theta_0)P_\ell(\cos\theta_0)\ell(\ell+1)m(m+1)g_{m\ell},&&\label{eq:coeffbl}
\end{eqnarray}
where $g_{\ell m} = g_{m\ell}$ and equals 
\begin{equation}
g_{m\ell} = \sum\limits_{\nu_n}\frac{\frac{d\nu_n}{d\theta_0}\frac{2\nu_n+1)}{\nu_n(\nu_n+1)}}{[\nu_n(\nu_n+1)-m(m+1)][\nu_n(\nu_n+1)-\ell(\ell+1)]}.
\label{eq:gml}\end{equation}
The matrix entries $g_{m\ell}$ are seen to depend only on the apex-angle $\theta_0$ of the antenna. This is in contrast to the version utilized in \cite[eqn. (31)]{janaswamy2022} and in all previous formulations of biconical antenna, where the matrix entries are combined functions of frequency and angle. However, (\ref{eq:coeffbl}) still suffers from the coefficients $b_\ell(k)$ being implicitly dependent on the variables $k, \theta_0$, and $L$. This implicit dependency has previously prevented the determination of explicit analytical forms of the temporal characteristics of the antenna. A change of variable
\begin{equation}
 j^{-\ell} \left[\widehat{H}^{(2)}_\ell(kL) + j\widehat{H}_\ell^{(2)\prime}(kL)\right]b_\ell(k) = \frac{2e^{-jkL}}{jkL}c_\ell(k)\label{eq:coeffcl}
\end{equation}
results in the following equation for the new coefficients $c_m(k)$, $m=1,3,\ldots$
\begin{eqnarray}
\frac{c_m(k)}{(2m+1)}&=& \frac{P_m(\cos\theta_0)}{m(m+1)} +\sin\theta_0P_m(\cos\theta_0)\times\nonumber\\
&&\sum\limits_{\ell=1,3,\ldots}^\infty c_\ell(k) P_\ell(\cos\theta_0)\ell(\ell+1)g_{m\ell}.\label{eq:cellEqn}
\end{eqnarray}
It is evident from (\ref{eq:cellEqn}) that the coefficients $c_m(k)$ depend only on the cone angle $\theta_0$. In particular, they are independent of frequency and, thereby, the argument $k$ can be safely dropped. One can also infer from (\ref{eq:coeffcl}) that the explicit frequency dependence of the coefficients $b_\ell(k)$ is governed entirely by the factors excluding $c_\ell$. This decomposition is a key advantage of our method. 

From the continuity of the combined field $E_\theta+\eta H_\phi$ at $r=a$ one further gets
\begin{equation}
2C_0\sum\limits_{\ell=1,3,\ldots}^\infty c_\ell P_\ell(\cos\theta_0)=: \sum\limits_{\ell=1,3,\ldots}^\infty w_\ell(\theta_0)= 1,\label{eq:check1} 
\end{equation}  
an identity for the normalized weights, $w_\ell(\theta_0)=2C_0c_\ell P_\ell(\cos\theta_0)$, which serves as a numerical check for the accuracy in obtaining the coefficients. 

It is possible to combine (\ref{eq:cellEqn}) and (\ref{eq:check1}) to result in a linear system for the coefficients $c_m$ that has a better condition number than (\ref{eq:cellEqn}). The result is
\begin{equation}
\frac{m(m+1)}{2m+1}c_m=(1+\gamma_m)P_m(\cos\theta_0) +\sum\limits_{\ell=1,3,\ldots}^\infty G_{m\ell}c_\ell,\label{eq:cell2Eqn}
\end{equation}
where
\begin{eqnarray}
G_{m\ell} &=& P_\ell(\cos\theta_0)P_m(\cos\theta_0)\biggl[\ell(\ell+1)m(m+1)\sin\theta_0g_{m\ell}\nonumber\\
&& - 2C_0\gamma_m\biggr],\label{eq:Gml}
\end{eqnarray}
and 
\begin{equation}
\gamma_m = [m(m+1)]^2g_{mm}\sin\theta_0/2C_0.\label{eq:gamm}
\end{equation}
Note that $G_{mm} = 0$. 

From the continuity of $E_\theta-j\eta H_\phi$, an expression for the reflection coefficient in terms of the normalized frequency $\zeta = kL$ is obtained as
\begin{equation}
e^{j(2\zeta+\pi)}\Gamma_{\rm in}(\zeta) = \sum\limits_{\ell=1,3,\ldots}^\infty w_\ell(\theta_0)\frac{\widehat{H}_\ell^{(2)}(\zeta)-j\widehat{H}^{(2)\prime}_\ell(\zeta)}{\widehat{H}_\ell^{(2)}(\zeta)+j\widehat{H}^{(2)\prime}_\ell(\zeta)}.\label{eq:Gamin}
\end{equation}
The dependence of the reflection coefficient on explicit functions of frequency as given in (\ref{eq:Gamin}) in another key advantage of the present method. Using the small argument approximation of the Hankel function $\widehat{H}^{(2)}_\ell(\zeta)\sim 2j(2\ell-1)!/(2\zeta)^\ell(\ell-1)!$, $|\zeta|\ll \ell$ and the large argument approximation $\widehat{H}^{(2)}_\ell(\zeta)\sim j^{\ell+1}e^{-j\zeta},\ \abs{\zeta}\gg\ell$, it is straightforward to make the inference 
\begin{equation}
\Gamma_{\rm in}(\zeta\sim 0) = 1+O(\zeta);\quad \Gamma_{\rm in}(\infty) = 0,\label{eq:Gamlmts}
\end{equation}
in view of (\ref{eq:check1}). 



\ifCLASSOPTIONcaptionsoff
  \newpage
\fi



%
\bibliographystyle{IEEEtran}
\bibliography{BiconRef.bib}

%





\end{document}